\documentclass[a4paper,11pt]{article}
\usepackage{amssymb,graphicx,amsmath,color}

\renewcommand{\theequation}{\thesection.\arabic{equation}}
\input{tcilatex}

\newcommand{\bc}{\begin{center}}
\newcommand{\ec}{\end{center}}
\def\ba#1{\begin{array}{#1}\displaystyle}
\newcommand{\ea}{\end{array}}

\newcommand{\beq}{\begin{equation}}
\newcommand{\eeq}{\end{equation}}
\newcommand{\beqa}{\begin{eqnarray}}
\newcommand{\eeqa}{\end{eqnarray}}

\newcommand{\bi}{\begin{itemize}}
\newcommand{\ei}{\end{itemize}}

\def\b#1{\bar{#1}}
\def\frc#1#2{\frac{#1}{#2}}

\newcommand{\bra}{\langle}
\newcommand{\ket}{\rangle}

\newcommand{\Tr}{{\rm Tr}}

\begin{document}

\setcounter{page}{0} \topmargin0pt \oddsidemargin0mm \renewcommand{%
\thefootnote}{\fnsymbol{footnote}} \newpage \setcounter{page}{0}
\begin{titlepage}
\vspace{0.2cm}
\begin{center}
{\Large {\bf Higher particle form factors of branch point twist
fields in integrable quantum field theories}}

\vspace{0.8cm} {\large \text{Olalla A.~Castro-Alvaredo and
Emanuele Levi}}

\vspace{0.2cm}
Centre for Mathematical Science, City University London, \\
Northampton Square, London EC1V 0HB, UK
\end{center}
\vspace{1cm} In this paper we compute higher particle form factors
of branch point twist fields. These fields were first described
 in the context of massive 1+1-dimensional integrable quantum
field theories and their correlation functions are related to the bi-partite
entanglement entropy. We find analytic expressions for some form
factors and check those expressions for consistency, mainly by
evaluating the conformal dimension of the corresponding twist
field in the underlying conformal field theory. We find that
solutions to the form factor equations are not unique so that
various techniques need to be used to identify those corresponding
to the branch point twist field we are interested in. The models
for which we carry out our study are characterized by staircase
patterns of various physical quantities as functions of the energy
scale. As the latter is varied, the $\beta$-function associated to
   these theories comes close to vanishing at several  points between
   the deep infrared and deep ultraviolet regimes. In other words, renormalisation
    group flows approach the vicinity of various critical points before ultimately reaching
    the ultraviolet fixed point. This feature
     provides an optimal way of checking the consistency of higher particle
     form factor solutions, as the changes on the conformal dimension of the
      twist field at various energy scales can only be accounted for
       by considering higher particle form factor contributions to the
       expansion of certain correlation functions.
 \vfill{
\hspace*{-9mm}
\begin{tabular}{l}
\rule{6 cm}{0.05 mm}\\
 \text{o.castro-alvaredo@city.ac.uk}\\
 \text{emanuele.levi.1@city.ac.uk}\\
\end{tabular}}

\renewcommand{\thefootnote}{\arabic{footnote}}
\setcounter{footnote}{0}

\end{titlepage}
\newpage
\section{Introduction}
Entanglement is the most distinct and bizarre of all quantum
phenomena. The idea that the quantum states of objects so wide
apart from each other that they are not even causally connected
can be entangled seems counterintuitive and has been hotly debated
\cite{epr}. Although the interpretation of quantum mechanics and
quantum entanglement continues to be discussed, the reality of
entanglement as a physical phenomenon was finally established
through the experiments of Alain Aspect and his collaborators
\cite{aspect}. In addition, in recent years, entanglement has
started to reveal its powerful practical applications, specially
in the context of quantum computation, quantum cryptology and
quantum teleportation (see e.g. \cite{quantum1,quantum2}).

From a theoretical point of view entanglement is intimately linked
to the structure of quantum states so that developing methods to
``measure" entanglement is a very efficient way to learn more
about the fundamental properties of quantum systems. In that
sense, many different theoretical measures of entanglement have
been proposed in the literature
\cite{bennet,Osterloh,Osborne,Barnum,Verstraete} of which
the \emph{entanglement entropy} \cite{bennet} is just an example. In recent
years much work has been carried out to compute the entanglement
entropy of extended quantum systems with many degrees of freedom,
such as quantum spin chains
\cite{Eisert,Latorre1,Latorre2,Latorre3,Jin,Lambert,KeatingM05,Weston,
ravanini,Ercolessi:2010eb,permutation} and quantum field theories
\cite{HolzheyLW94,Calabrese:2004eu,Calabrese:2005in,entropy,review}.

Consider a quantum system, with Hilbert space ${\cal H} = {\cal
H}_A\otimes{\cal H}_B$, in a pure state $|\psi\rangle$. The
bi-partite entanglement entropy $S_A$ is the von Neumann entropy
\cite{von} associated to the reduced density matrix of the
subsystem $A$, $\rho_A$, defined as
 \beq
    \rho_A = \Tr_{{\cal H}_{B}}(|\psi\ket\bra \psi|)\,, \qquad
    S_A = -\Tr_{{\cal H}_A} (\rho_A \log(\rho_A))~.
    \label{entropy}
\eeq One way of interpreting $S_A$ is to understand it as a
measurement of the entanglement of the quantum state of subsystem
$A$ when the latter is considered in isolation (ignoring the
existence of subsystem $B$).

In a series of recent works involving one of the present authors
\cite{entropy,other,nexttonext,review} a new approach to the
computation of (\ref{entropy}) for 1+1-dimensional integrable
quantum field theories (QFT) has been proposed and developed. This
approach takes as starting point the ``replica trick". This
consists of replacing the theory under scrutiny by a new model
consisting of $n$ non-interacting copies or ``replicas" of the
original theory. Although this might seem to lead to an
unnecessary complication of the problem, it does in fact simplify
it by producing a new theory which possesses an extra symmetry under cyclic permutations of the $n$-copies of the model.
Associated to this symmetry there exists a special class of twist
fields $\mathcal{T}$ and $\tilde{\mathcal{T}}=\mathcal{T}^\dagger$
which have been named \emph{branch point twist fields}. The key
result is
\begin{equation}
    \text{Tr}_{{\cal H}_A}\rho_A^n\sim \langle \mathcal{T}(r)
   \tilde{ \mathcal{T}}(0)\rangle, \label{ide}
\end{equation}
that is, the trace of the density matrix of the replica theory is
proportional to the two-point function of twist fields. $r$ is the
size of subsystem $A$ in the quantum field theory. Therefore, once
the trace has been computed, the von Neumann entropy follows from
the identity,
 \beq\label{SA}
    S_A = -\lim_{n\to 1} \frc{d}{d n} \Tr_{{\cal H}_A} \rho_A^n~.
\eeq This identity does however involve a highly non-trivial
analytic continuation of the function $\Tr_{{\cal H}_A} \rho_A^n$
to positive non-integer values of $n$ which makes the limit above
extremely hard to compute \cite{entropy,other,nexttonext}.
It is sometimes more convenient
to compute another type of entanglement entropy known as
 R\'enyi entropy \cite{renyi} which
is given by
\begin{equation}
    S_A^{\text{R\'enyi}}(n)=\frac{\log\left(\Tr_{{\cal A}}\left( \rho_A^n
    \right)\right)}{1-n},\label{renyi}
\end{equation}
and whose $n \rightarrow 1$ limit gives (\ref{entropy}) once more.

In the current work we wish to take the identity (\ref{ide}) as
our main motivation to investigate the form factors of the twist
field $\mathcal{T}$. As is well known, the correlation function in
(\ref{ide}) can be expanded in terms of form factors of the fields
involved. Therefore knowing the form factors allows us in
principle to extract both the von Neumann and R\'enyi entropies of
any integrable QFT under consideration.

This paper is organised as follows: In section 2 we summarise the form
factor programme for branch point twist fields \cite{entropy}.
In section 3 we introduce the $\Delta$-sum rule \cite{DSC} and
the two models which we want to study in detail: the roaming trajectories
(RT) model and the $SU(3)_2$-homogenous sine-Gordon (HSG) model. In section 4 we describe in detail the construction of higher particle form factors for the RT-model. We place special emphasis on the existence of multiple solutions to the form factor equations. We successfully apply the cluster decomposition property to identify those that must correspond to the branch point twist field which is related to the bi-partite entanglement entropy of the model. In section 5 we carry out a similar analysis for the $SU(3)_2$-HSG model concentrating less on the non-uniqueness of solutions and more on the extra challenges posed by the more complex particle spectrum of the model. In section 6 we provide numerical results for the ultraviolet conformal dimension of the twist field in the RT- and HSG-model which are fully consistent with the theoretical predictions. We present our conclusions in section 7.

\section{Form factors of branch point twist fields}
In this paper we will be concerned with the computation
 of matrix elements of the branch point twist field $\mathcal{T}$
  for particular models. Once the form factors are known they may
  be used in the expansion of any correlation functions involving
   the twist field. For example, the two-point function
$\langle \mathcal{T}(0)
  \mathcal{O}(r)\rangle$ can be written as
\begin{eqnarray}
\langle \mathcal{T}(r) \mathcal{O}(0)\rangle &=&
\sum_{k=0}^\infty\frac{1}{k!} \sum_{\mu_1,\ldots,\mu_k=1}^n \left(
\prod_{j=1}^k \int \limits_{-\infty }^{\infty } \frac{d\theta
_{j}}{(2\pi )}\right) F_{k}^{\mathcal{T} |\mu_1 \ldots
\mu_k}(\theta_1,\ldots, \theta_{k};n) \nonumber \\
&& \times \left(F_{k}^{\mathcal{O} |\mu_1 \ldots
\mu_k}(\theta_1,\ldots, \theta_{k}) \right)^* \,e^{-r
\sum\limits_{j=1}^k m_{\mu_i}\cosh\theta_j},\label{ent1}
\end{eqnarray}
where $\mathcal{O}$ represents a generic local field of the theory.
Here $F_{k}^{\mathcal{T} |\mu_1 \ldots \mu_k}(\theta_1,\ldots,
\theta_{k},n)$ is the $k$-particle form factor of the field
$\mathcal{T}$ (similarly for the other field) which is defined as
\begin{equation}
F_{k}^{\mathcal{T}|\mu _{1}\ldots \mu _{k}}(\theta _{1},\ldots
,\theta _{k};n):=\left\langle
0|\mathcal{T}(0)|\theta_1,\ldots,\theta_k\right\rangle_{\mu_1,\ldots,\mu_k}^{\text{in}}
~,\label{ff}
\end{equation}
where $|0\rangle$ represents the vacuum state and
$|\theta_1,\ldots,\theta_k\rangle_{\mu_1,\ldots,\mu_k}^{\text{in}}$
are the physical ``in'' asymptotic states of massive QFT. They
carry indices $\mu_i$, which are double indices of the form
\begin{equation}\label{index}
   \mu_i:=(\alpha_i,c_i),
\end{equation}
where $\alpha_i$ labels the particle species and $c_i$ labels the copy number within the replica theory.
The mass of the corresponding particle is denoted by $m_{\mu_i}$ and its energy and momentum are parameterized
by the real parameter $\theta_i$, called the rapidity. The form factors also depend on $n$, the number of replicas of the
model whose entropy we want to investigate. In particular, for $n=1$ the twist field should reduce to the identity field so that all form factors, except the
vacuum expectation value should vanish.

For most QFTs the form factors defined above are only accessible
perturbatively. However, one of the remarkable advantages of
integrable QFTs is that integrability  is constraining enough as
to almost fix the functions (\ref{ff}) completely. This feature
was realized several decades ago \cite{Weisz,KW,smirnovbook} when
it was shown that form factors such as (\ref{ff}) may be
systematically obtained as the solutions to a set of consistency
equations which only require the knowledge of the scattering (S)
matrix of the theory under consideration as input. This solution
process is sometimes referred to as the \emph{form factor
programme}. This fact has triggered an enormous amount of work in
computing form factors in a multitude of models of integrable QFT
(see \cite{Karowski:1979dk,Essler:2004ht,musbook} for reviews).

The special nature of the twist fields $\mathcal{T}$ and
$\tilde{\mathcal{T}}$ has however made it necessary to rethink the
form factor programme described above from first principles in
order to adapt it to replica theories \cite{entropy}. In order to make expressions more compact, we will be dropping the
explicit $n$-dependence of the form factors defined by (\ref{ff}).
If we
consider integrable QFTs without backscattering or bound states
(theories with backscattering and bound states have been
considered in \cite{nexttonext} and \cite{entropy}, respectively),
then the new form factor equations are given by
\begin{eqnarray}
  F_{k}^{\mathcal{T}|\ldots \mu_i  \mu_{i+1} \ldots }(\ldots,\theta_i, \theta_{i+1}, \ldots ) &=&
  S_{\mu_i \mu_{i+1}}^{(n)}(\theta_{i\,i+1})
  F_{k}^{\mathcal{T}|\ldots \mu_{i+1}  \mu_{i} \ldots}(\ldots,\theta_{i+1}, \theta_i,  \ldots ),
  \label{1}\\
 F_{k}^{\mathcal{T}|\mu_1 \mu_2 \ldots \mu_k}(\theta_1+2 \pi i, \ldots,
\theta_k) &=&
  F_{k}^{\mathcal{T}| \mu_2 \ldots \mu_n \hat{\mu}_1}(\theta_2, \ldots, \theta_{k},
  \theta_1),\label{2}
  \end{eqnarray}
  where $\theta_{ij}=\theta_i-\theta_j$ and $\hat{\mu}_i=(\alpha_i,c_i+1)$.
  The function $S_{\mu_i \mu_{i+1}}^{(n)}(\theta_{i\,i+1})$ is the two
  particle  S-matrix of the replica theory, defined as
\begin{equation}
  S_{\mu_i \mu_{i+1}}^{(n)}(\theta)
  =\left\{\begin{array}{l} 1,\qquad\qquad\,\,\,\,\,
     \text{iff}\quad c_i\neq c_{i+1} \\
  S_{\alpha_i \alpha_{i+1}}(\theta), \quad \text{iff}\quad c_i=c_{i+1}
                                          \end{array}
                                        \right.
\end{equation}
and $ S_{\alpha_i \alpha_{i+1}}(\theta)$ represents the $S$-matrix of the
original (non-replica) model. Equations (\ref{1}) is the same for form factors of other local fields, whereas equation (\ref{2}) is slightly different, in that the index $\hat{\mu}_i$ is involved. This difference is responsible for the fact that, on top of the standard kinematic poles at
$\theta_{ij}=i\pi$, the form factors have extra poles at $\theta_{ij}=i\pi(2n-1)$ in the extended physical strip $\Im(\theta_{ij})\in [0,2\pi n)$.
Thus we have two kinematic residue equations, which provide expressions for the residues at those poles
  \begin{eqnarray}
 \begin{array}{l}
\\
  \text{Res}  \\
 {\footnotesize \bar{\theta}_{0}={\theta}_{0}}
\end{array}\!\!\!\!
 F_{k+2}^{\mathcal{T}|\b{\mu} \mu  \mu_1 \ldots \mu_k}(\bar{\theta}_0+i\pi,{\theta}_{0}, \theta_1 \ldots, \theta_k)
  &=&
  i \,F_{k}^{\mathcal{T}| \mu_1 \ldots \mu_k}(\theta_1, \ldots,\theta_k), \label{3}
  \\
\begin{array}{l}
\\
  \text{Res}  \\
 {\footnotesize \bar{\theta}_{0}={\theta}_{0}}
\end{array}\!\!\!\!
 F_{k+2}^{\mathcal{T}|\b\mu \hat{\mu } \mu_1 \ldots \mu_k}(\bar{\theta}_0+i\pi,{\theta}_{0}, \theta_1 \ldots, \theta_k)
  &=&-i\prod_{i=1}^{k} S_{\hat{\mu}\mu_i}^{(n)}(\theta_{0i})
  F_{k}^{\mathcal{T}| \mu_1 \ldots \mu_k}(\theta_1, \ldots,\theta_k).\label{kre}
\end{eqnarray}
Here $\b\mu=(\bar{\alpha},c)$ is the anti-particle of $\mu$. The equations (\ref{3}) and
(\ref{kre}) are in fact not independent from each other, as
solutions to (\ref{kre}) may be obtained from the solutions to
(\ref{3}) by employing the first two form factor equations. This
is the reason why later on we will concentrate on solving
(\ref{3}).

\subsection{Minimal form factors}
In order to solve equations (\ref{1})-(\ref{kre}) it is common practise to adopt a recursive approach whereby one starts by computing the two-particle form factors (the zero-particle and one-particle form factors are constant for spinless fields, with $F_0^{\mathcal{T}}=\langle 0|\mathcal{T}|0 \rangle :=\langle \mathcal{T}\rangle $) and then uses equations (\ref{3})-(\ref{kre}) to obtain higher particle form factors iteratively. A minimal solution (e.g. with no poles on the physical sheet) can be obtained by solving the two particle versions of equations (\ref{1}) and (\ref{2}), that is
\begin{equation}
    F_{\text{min}}^{\mathcal{T}| \mu_1 \mu_2}(\theta)=
    S_{\mu_1 \mu_2}(\theta)F_{\text{min}}^{\mathcal{T}|\mu_2 \mu_1}(-\theta)
    =F_{\text{min}}^{\mathcal{T}|\mu_2 \mu_1}(2\pi i n-\theta), \qquad \forall
    \qquad \mu_1\,, \mu_2 \label{fmin}
\end{equation}
which, combined with the requirement that the two particle form factor must have simple poles at $\theta=i \pi$ and $\theta=i \pi(2n-1)$ and that the residue at the first pole must be
\begin{eqnarray}
 \begin{array}{l}
\\
  \text{Res}  \\
 {\footnotesize \theta=0}
\end{array}
 F_{2}^{\mathcal{T}|\b{\mu} \mu }(\theta+i\pi)
  &=&
  i F_0^{\mathcal{T}}:= i \langle \mathcal{T} \rangle, \label{ni3}
\end{eqnarray}
gives the general solution \cite{entropy}
\begin{equation}
F_{2}^{\mathcal{T}|\mu_1 \mu_2}(\theta)=\frac{ \langle
\mathcal{T}\rangle \sin\left(\frac{\pi}{n}\right)}{2 n
\sinh\left(\frac{ i\pi (2(c_1 -c_2)-1)+
\theta}{2n}\right)\sinh\left(\frac{i\pi (2(c_2-
c_1)-1)-\theta}{2n}\right)}
\frac{F_{\text{min}}^{\mathcal{T}|\mu_1
\mu_2}(\theta)}{F_{\text{min}}^{\mathcal{T}|\mu_1 \mu_2}(i
\pi)},\label{full}
\end{equation}
where $c_{1,2}$ represent the copy number associated to each of
the particles as defined in (\ref{index}).

\section{Introducing the models and the $\Delta$-sum rule}\label{deltasec}
In \cite{entropy} a general expression for the two-particle form
factors of the twist field was found that applies to all
integrable QFTs. This expression was then specialised to the Ising
and sinh-Gordon models and checked against the $\Delta$-sum rule
\cite{DSC}. In its original form, the $\Delta$-sum rule can be expressed as
\begin{equation}\label{delta}
    \Delta^{\mathcal{O}} =-\frac{1}{2 \langle \mathcal{O}\rangle}\int_{0}^{\infty} d r\, r \langle \Theta(r) \mathcal{O}(0)\rangle_c,
\end{equation}
where the subindex $c$ in the two-point function stands for
``connected",  meaning that the product of the vacuum expectation
values of the fields has been subtracted. $\Theta$ is the trace of
the energy momentum tensor of the model. Eq.~(\ref{delta})
provides an expression for the conformal dimension of a primary
field. The remarkable fact is that the dimension is given in
terms of a two point function involving the field $\mathcal{O}$
which represents the counterpart of the primary field in the
perturbed (massive) model. Therefore (\ref{delta}) allows us to
extract information about the underlying CFT from the two-point function of a massive theory.

A slightly more general version of (\ref{delta}) was employed in
\cite{CastroAlvaredo:2000ag}
\begin{equation}\label{delta2}
    \Delta^{\mathcal{O}}(r_0) =-\frac{1}{2 \langle \mathcal{O}\rangle}\int_{r_0}^{\infty} d r \,r \langle \Theta(r) \mathcal{O}(0)\rangle_c,
\end{equation}
so that taking $r_0=0$ we recover (\ref{delta}) whereas for larger values of $r_0$ we are now able to trace changes in the value of $\Delta^{\mathcal{T}}(r_0)$ along the renormalisation group (RG) flow, that is as we move from low energies or $r_0$ large to high energies or $r_0=0$. Observing such intermediate behaviour is particularly interesting for models where the RG-flows approach the vicinity of more than one critical point, as the ones we will consider below.

Taking $\mathcal{O}=\mathcal{T}$, employing (\ref{ent1}) and
performing the integration in $r$ (\ref{delta2}) becomes
 \begin{eqnarray}
\label{expansion1}
\Delta^{\mathcal{T}}(r_{0}) &=&-\frac{1}{%
2\left\langle \mathcal{T}\right\rangle }\sum_{k=1}^{\infty
}\sum_{\mu _{1}\ldots \mu _{k}}\int\limits_{-\infty }^{\infty
}\ldots
\int\limits_{-\infty }^{\infty }\frac{d\theta _{1}\ldots d\theta _{k}}{%
k!(2\pi )^{k}} \frac{\left( 1+r_{0}E \right)e^{-r_{0}E}}{2E^{2}}
\nonumber \\
&&\times F_{k}^{\Theta |\mu _{1}\ldots \mu _{k}}(\theta
_{1},\ldots ,\theta _{k})\,\left( F_{k}^{\mathcal{T}|\mu
_{1}\ldots \mu _{k}}(\theta _{1},\ldots ,\theta _{k})\,\right)
^{*}\,\,\,, \label{exexpansion1}
\end{eqnarray}
where $E$ stands for the sum of the on-shell energies $E=\sum_{i=1}^{k}m_{\mu_{i}}\cos(\theta_{i})$ \cite{CastroAlvaredo:2000ag}.
In those models for which the form factors of $\Theta$ and $\mathcal{T}$ are known,
one can in principle identify the conformal dimension
$\Delta^\mathcal{T}$ by computing (\ref{exexpansion1}) and taking $r_{0}=0$.
This dimension is known a priori to be
\begin{equation}
    \Delta^\mathcal{T}(0):=\Delta^\mathcal{T}=\frac{c}{24}\left(n-\frac{1}{n}\right),
    \label{known1}
\end{equation}
so that evaluating (\ref{delta}) effectively allows for a consistency check of the form factors of $\mathcal{T}$.
Such a check was carried out successfully in \cite{entropy} for the Ising and sinh-Gordon models.
There only the two-particle contribution to the expansion of the two-point function was considered and very good agreement with
the predicted value was reached (the agreement was exact for the Ising model).
The reason why the two-particle approximation works so well is that the expansion (\ref{ent1}) is in fact rapidly convergent as a function of the particle numbers involved in the form factors.
Therefore, in general the two-particle contribution is the more major and for many models other contributions are almost negligible.

Although this is a very useful feature which has been widely exploited for integrable models, it is inconvenient if one wants to use (\ref{exexpansion1}) to test higher particle form factor solutions.
There is however a way to overcome this obstacle, that is, by considering models for which the two particle contribution is far from providing a good picture of the ultraviolet behaviour of the theory. In the upcoming subsections we will describe two models which have precisely this feature: the roaming trajectories model and the $SU(3)_2$-homogeneous sine-Gordon model.
\subsection{The roaming trajectories model}\label{rtmodel}
The first theory we want to investigate here is the \emph{roaming
trajectories (RT) model} \cite{roaming}. This is a model with a
single particle spectrum and no bound states which is closely
related to the sinh-Gordon model. The model is characterized by
the two-particle S-matrix,
\begin{equation}\label{roaming}
    S(\theta)=\tanh\frac{1}{2}\left(\theta-\theta_0- \frac{i\pi}{2}\right)
    \tanh\frac{1}{2}\left(\theta+\theta_0- \frac{i\pi}{2}\right),
    \quad\theta_0 \in \mathbb{R}.
\end{equation}
On the other hand, the sinh-Gordon S-matrix \cite{toda1,toda2} is
given by
\begin{equation}\label{shs}
    S(\theta)=\frac{\tanh\frac{1}{2}\left(\theta- \frac{i\pi B}{2}\right)}{
    \tanh\frac{1}{2}\left(\theta+ \frac{i\pi B}{2}\right)}, \quad B \in
    [0,2].
\end{equation}
It is easy to see that the S-matrix (\ref{roaming}) can be
obtained from (\ref{shs}) by the replacement
\begin{equation}
B \rightarrow 1-\frac{2 i \theta_0}{\pi}. \label{rep}
\end{equation}
This relationship implies in particular that computing the form
factors of the sinh-Gordon model and setting $B$ to the value
(\ref{rep}) gives the form factors of the RT-model.

The roaming trajectories the model's name refers to emerged in the
computation of the effective central charge $c_{\text{eff}}(r)$
within the thermodynamic Bethe ansatz approach \cite{tba1,tba2}
carried out in \cite{roaming}. For massive QFTs it is expected
that the function $c_{\text{eff}}(r)$ ``flows" from the value zero
in the infrared (large $r$) to a finite value in the ultraviolet
(small $r$). For many theories, including the sinh-Gordon model,
the constant value reached as $r\rightarrow 0$ is the central
charge of the underlying conformal field theory associated to the
model. In this case, that theory is the free massless boson, a
conformal field theory with central charge $c=1$. Therefore, in
the sinh-Gordon model, the function $c_{\text{eff}}(r)$ flows from
the value zero to the value 1 as $r$ decreases.

Crucially, when the same
function $c_{\text{eff}}(r)$ is computed for the RT-model it shows a
very different behaviour. It still flows from the value 0 to the
value 1, but it does so by ``visiting" infinitely many
intermediate values of $c$ giving rise to a staircase (or
roaming) pattern. The values of $c$ that are visited
correspond exactly to the central charges of the unitary minimal
models of conformal field theory
\begin{equation}\label{cc}
    c_p=1-\frac{6}{p(p+1)},\quad \text{with} \quad p=3,4,5\ldots
\end{equation}
Another observation made in \cite{roaming} is that the size of the
intermediate plateaux that the function $c_{\text{eff}}(r)$
develops at the values (\ref{cc}) is determined by the value of
$\theta_0$.  For
$\theta_0=0$ there is a single plateaux at $c=1$, thus the usual
sinh-Gordon behaviour is recovered, whereas the plateaux at
(\ref{cc}) become more prominent as $\theta_0$ is increased. In
the limit $\theta_0\rightarrow \infty $ a single plateaux at
$c=\frac{1}{2}$ remains which reflects the fact that the
$S$-matrix (\ref{roaming}) becomes -1 in this limit, hence the
model reduces to the Ising field theory. This interesting limit
behaviour was studied in \cite{Ahn:1993dm} within the form factor
approach.
\subsection{The $SU(3)_2$-Homogeneous sine-Gordon model}\label{sgmodel}
The second model we want to study is the $SU(3)_2$-Homogeneous
sine-Gordon (HSG) model. The model is just one of the simplest
representatives of a large class of theories first named in \cite{ntft} whose spectrum
\cite{ntft,hsg,FernandezPousa:1997iu}, S-matrix \cite{smatrix},
form factors
\cite{CastroAlvaredo:2000em,CastroAlvaredo:2000nk,CastroAlvaredo:2000ag,CastroAlvaredo:2000nr}
and thermodynamic properties
\cite{CastroAlvaredo:1999em,CastroAlvaredo:2002nv,Dorey:2004qc}
have been extensively investigated over the last two decades. The
HSG-models are very interesting theories, as they include a number
of distinct features rarely found for integrable  models: they
posses both unstable particles and bound states in their spectrum
and their $S$-matrices are generally non-parity invariant, that is
$S_{ab}(\theta)\neq S_{ba}(\theta)$ for $a\neq b$. In particular,
the $SU(3)_2$-HSG model contains two particles, which we will label
as $+$ and $-$. They are self-conjugated and interact with each
other by means of the following S-matrix
\begin{equation}
    S_{\pm \pm}(\theta)=-1, \qquad\text{and} \qquad
     S_{\pm \mp}(\theta)=\pm \tanh\frac{1}{2}\left(\theta \pm \sigma -
     \frac{i\pi}{2}\right).\label{hsg}
\end{equation}
Thus particles of the same species interact with each other as
free fermions, whereas particles of different species
interact by means of parity-breaking S-matrix which depends on a
free parameter $\sigma$. These S-matrix amplitudes have a pole in
the unphysical sheet (that is $\Im(\theta)\in (-\pi,0) $),
with real part given by $\pm \sigma$. Such
type of poles are a signature of the presence of unstable
particles in the spectrum.

The scattering picture is that particles $+$ and $-$
interact with each other by creating an unstable particle, whose
mass and decay width depend on the parameter $\sigma$
through Breit-Wigner's formula \cite{bw}. More precisely, for $|\sigma|$ large, the mass
of the unstable particle can be approximated by $m e^{|\sigma|/2}$, where $m$ is the mass
of the stable particles \cite{CastroAlvaredo:2000ag}. Therefore, the limit $\sigma \rightarrow \infty$ corresponds to an infinitely massive unstable particle, that is a particle that can not be formed at any finite energy scales. At the level of the S-matrix we find that $\lim_{\sigma\rightarrow\infty} S_{\pm,\mp}(\theta)=1$, that is, the model reduces to two non-interacting copies of the Ising field theory.
This property is very useful as a consistency check in form factor calculations. It
implies that when $\sigma \rightarrow \infty$ the form factors of
any field should reduce to those of the Ising model, which are generally known.

As for the RT-model described before, the effective central charge
of the $SU(3)_2$-HSG model also exhibits a staircase pattern,
albeit with only two steps (at most) \cite{CastroAlvaredo:1999em}.
The same structure was found for Zamolodchikov's $c$-function and
the conformal dimensions of certain local fields
\cite{Zamc,CastroAlvaredo:2000ag}. In this case the appearance of
steps is directly related to the presence of the unstable particle
and its mass. There is only one step if $\sigma=0$ in which case
the unstable particle's mass is of the same order as that of the
stable particles and a second step emerges if $\sigma\neq 0$ whose
onset and length are related to the precise value of $\sigma$. All
these features have been analysed in detail in
\cite{CastroAlvaredo:1999em,CastroAlvaredo:2000ag}. In section 6
we will see that the conformal dimension of the twist field
(\ref{exexpansion1}) is no exception to this behaviour.

\section{Twist field form factors for the RT-model}
We will start our analysis by considering the simplest of the two models described above,
in terms of its particle spectrum. The two-particle minimal form
factor of the sinh-Gordon model
\begin{equation}
 F_{\text{min}}^{\mathcal{T}|11}(\theta)=\exp\left[-2
 \int_{0}^{\infty} \frac{dt \sinh \frac{t B}{4} \sinh  \frac{t(2-B)}{4}}{t \sinh(n t)
 \cosh \frac{t}{2} } \cosh t\left(n+\frac{i\theta}{\pi}
    \right)\right],\label{int}
\end{equation}
was first obtained in \cite{entropy} and can be easily rewritten as an infinite product of
ratios of Gamma functions. The explicit expression can be also
found in \cite{entropy}.

It is natural to make the following ansatz,
\begin{equation}
 F_{k}^{\mathcal{T}}(x_1,\ldots,x_k)=H_k Q_k(x_{1},...,x_{k}) \prod_{i<j}^{k}
\frac{F_{\text{min}}^{\mathcal{T}|11}(\frac{x_i}{x_j})}{(x_i-\alpha
x_j)(x_j-\alpha x_i)},\label{ansatz2}
\end{equation}
where we have introduced the new variables
$x_i=e^{\frac{\theta_i}{n}}$ and $\alpha=e^{\frac{i\pi}{n}}$ so
that, for example
\begin{equation}
   F_{\text{min}}^{\mathcal{T}|11}(\theta_i-\theta_j)\equiv
   F_{\text{min}}^{\mathcal{T}|11}(\frac{x_i}{x_j}),
\end{equation}
A similar ansatz was already used in \cite{niedermaier} in a different context.
We use the simplified notation $ F_{k}^{\mathcal{T}}(x_1,\ldots,x_k)$ to represent
the $k$-particle form factor of particles all of which live in the same copy
of the model. The functions
$Q_{k}(x_{1},...,x_{k})$ are symmetric in all variables and have
no poles on the physical sheet. $H_k$ are rapidity independent.

The ansatz (\ref{ansatz2}) is reminiscent of the solution procedure
that is traditionally used in the original form factor programme
(see e.g. \cite{FMS} where the sinh-Gordon model was studied).
This ansatz is useful as it isolates the pole structure of the
form factors in the product. Provided that $Q_{k}(x_{1},...,x_{k})$ are analytic functions,
symmetric in all variables, then it automatically satisfies equations
(\ref{1}) and (\ref{2}). For $k=0$, the condition
(\ref{ni3}) implies the normalization $H_0 = \langle
\mathcal{T}\rangle$ and $Q_0=1$.

Once the ansatz (\ref{ansatz2}) has been made it remains to
identify the functions $Q_k(x_1,\ldots,x_k)$ and the constants
$H_k$. In the sinh-Gordon model symmetry considerations imply that
only even particle form factors are non-vanishing, so that our
first new results would correspond to the $k=4$ case and $k$ will
always be an even number. We therefore turn to solving equation
(\ref{3}), which we can now rewrite as
\begin{equation}
\lim_{\bar{\theta}_0 \to\theta_0}(\bar{\theta}_0
-\theta_0)F_{k+2}(\alpha x_{0},x_{0},x_{1},\ldots,x_{k})= i
F_{k}(x_{1},\ldots,x_{k}), \label{equ:rec1}
\end{equation}
where $x_0=e^{\frac{\theta_0}{n}}$.

In order to turn the equation (\ref{equ:rec1}) into an equation
for the functions $Q_k(x_1,\ldots,x_k)$ and the constants $H_k$
the following identity will be needed, \beq
F_{\text{min}}^{\mathcal{T}|11}(\frac{\alpha
x_0}{x_i})F_{\text{min}}^{\mathcal{T}|11}(\frac{x_0}{x_i})=
\frac{(x_0-x_i)(\alpha x_0 -x_i)}{(\alpha \beta^{-1}x_0-x_i)(\beta
x_0 -x_i)}, \eeq where $\beta=e^{\frac{i \pi B}{2n}}$ and $B$ is
the coupling constant that appears in the sinh-Gordon $S$-matrix
(\ref{shs}). This identity can be easily derived from the Gamma
function representation of the minimal form factor \cite{entropy}.

Substituting the ansatz (\ref{ansatz}) into (\ref{equ:rec1}) and
simplifying we obtain
\begin{equation}
    H_{k+2}=\frac{2\sin{\frac{\pi}{n}}\,\alpha^{k+2}}
    {n F_{\text{min}}^{\mathcal{T}|11}(i\pi)} H_k\qquad \text{and} \qquad
Q_{k+2}(\alpha x_0,x_0,x_{1},\ldots,x_{k})=
 x_0^{2}P_{k}Q_{k}(x_{1},\ldots,x_{k}).
\label{qk}
\end{equation}
where
\begin{eqnarray}
 P_k &=& \prod_{a,b,c,d=1}^k
 (x_a-\alpha^2 x_0)(x_0-\alpha x_b)
 (x_c-\alpha \beta^{-1}x_0)(\beta x_0 -x_d) \\
&=&(-\alpha)^k
\sum^{k}_{a,b,c,d=0}(-\alpha^{2}x_0)^{k-a}(-\alpha^{-1}{x_0})^{k-b}
(-{\alpha}{\beta^{-1}}x_0)^{k-c}(-\beta x_0)^{k-d}\sigma_a^{(k)}
\sigma_b^{(k)} \sigma_c^{(k)} \sigma_d^{(k)},\nonumber
\end{eqnarray}
and $\sigma_i^{(k)}$ is the $i$-th elementary symmetric
polynomial on $k$ variables $x_1,\ldots,x_k$, which can be defined
by means of the generating function,
\begin{equation}\label{sigma}
   \sum_{i=0}^k x^{k-i} \sigma_i^{(k)}= \prod_{i=1}^k(x_i + x).
\end{equation}
The equation for $H_k$ can be easily solved to
\begin{equation}\label{solhk}
    H_k=\left(\frac{2\sin{\frac{\pi}{n}}\,\alpha^{2}}
    {n F_{\text{min}}^{\mathcal{T}|11}(i\pi)}\right)^{\frac{k}{2}} \alpha^{\frac{k}{2}(\frac{k}{2}-1)} \langle \mathcal{T}
    \rangle,
\end{equation}
whereas equations for the polynomials $Q_k(x_1,\ldots,x_k)$ will need to be solved on a case by
case basis. Unfortunately the solutions get very involved very quickly. There are three main reasons for this:
\begin{itemize}
  \item The degree of the polynomial in the denominator of (\ref{ansatz2}) is much higher than would be the case in the standard form factor programme. Since the twist field is spinless, the degree of such polynomial must equal the degree of the polynomial $Q_k(x_1,\ldots,x_k)$ and this means that its degree will be very high for relatively small values of $k$. As an example, for the RT-model  we will see later that the degree of $ Q_2(x_1,x_2)$ is just 2, but the degrees of $Q_4(x_1,x_2,x_3,x_4)$ and $Q_6(x_1,\ldots,x_6)$ are 12 and 30 respectively.
  \item The polynomial $P_k$ is a very complicated function in terms of elementary symmetric polynomials, which again complicates the solution procedure and makes it very difficult to identify any patterns as $k$ is increased.
\item The reduction properties of the elementary symmetric polynomials $\sigma_i^{(k)}$ are much more involved for the twist field than in the usual form factor programme. In general,
\begin{equation}
    \sigma_i^{(k+2)}=  \sigma_{i}^{(k)}+ (1+\alpha)x_0\sigma_{i-1}^{(k)}+ \alpha x_0^2\sigma_{i-2}^{(k)},
\end{equation}
where $\sigma_i^{(k+2)}$ is an elementary symmetric polynomial on the variables $\alpha x_0, x_0,x_1,\ldots,x_k$ and $\sigma_i^{(k)},\sigma_{i-1}^{(k)},\sigma_{i-2}^{(k)} $ are elementary symmetric polynomials in the variables $x_1,\ldots,x_k$. We will also adopt the conventions $ \sigma_{i}^{(k)}=0$ for $i<0$ and $ \sigma_{0}^{(k)}=1$.
The usual reduction properties are recovered for $n=1$ or $\alpha=-1$.
\end{itemize}
The polynomial
$Q_2(x_1,x_2)$ can be easily obtained by setting $k=0$ in
(\ref{qk}) which gives the equation,
\begin{equation}
Q_{2}(\alpha x_0,x_0)=
 x_0^{2},\label{eqq2}
\end{equation}
There are actually two combinations of elementary symmetric polynomials of two variables
$\sigma_1^{(2)}$ and $\sigma_2^{(2)}$ that correctly reduce to
the identity above. The most general solution is
\begin{equation}
    Q_2(x_1,x_2)= \alpha^{-1}\sigma_2^{(2)}+ \Omega_2 K_2(x_1,x_2), \label{sol2}
\end{equation}
with $\Omega_2 $ an arbitrary constant and
\begin{equation}\label{k2}
    K_2(x_1,x_2)=\alpha^{-1}\sigma_2^{(2)}-\left(\frac{\sigma_1^{(2)}}{1+\alpha}\right)^2,
\end{equation}
the kernel of equation  (\ref{eqq2}), that is the most general order 2 polynomial on the variables $x_1, x_2$ which solves
\begin{equation}\label{kernel}
    Q_{k+2}(\alpha x_0,x_0,x_1,\ldots,x_k)=0,
\end{equation}
with $k=0$. Substituting (\ref{sol2}) together with $H_2$ in (\ref{ansatz2})
it is easy to see that (\ref{full}) is only recovered for $\mu_1=\mu_2=c_1=c_2=1$ if we choose $\Omega_2=0$. Hence we have fixed the constant above and can now go on to compute the four particle form factor.

Solving now for $Q_4(x_1,x_2,x_3,x_4)$ we find that the most general solution
to (\ref{qk}) takes the form
\begin{eqnarray}
  Q_4(x_1,x_2,x_3,x_4)&=&\sigma_4\left[ \sigma_2^4+
   \gamma \sigma_2 (\sigma_3^2+ \sigma_1^2 \sigma_4 ) +
   \delta \sigma_1\sigma_2^2 \sigma_3 +\eta \sigma_1^2\sigma_3^2
   + \xi \sigma_2^2 \sigma_4\right. \nonumber \\
   &&\left. + \lambda  \sigma_1\sigma_3\sigma_4+ \rho \sigma_4^2 \right]+ \Omega_4 K_4(x_1,x_2,x_3,x_4),\label{q4}
\end{eqnarray}
where we have abbreviated $\sigma_i^{(4)}\equiv \sigma_i$. The
constants $\gamma, \delta,\eta,\xi,\lambda$ and $\rho$ are fixed
functions of $n$, whose explicit form is given in appendix A. The
function $K_4(x_1,x_2,x_3,x_4)$ is the most general order 12 polynomial on the variables $x_1, x_2, x_3$ and $x_4$
that solves the equation (\ref{kernel}) and $\Omega_4$ is an arbitrary constant. The function $K_4(x_1,x_2,x_3,x_4)$ has the form
\begin{eqnarray}
  K_4(x_1,x_2,x_3,x_4)&=& A \sigma_1^2\sigma_2^2 \sigma_3^2 + B (\sigma_2^3 \sigma_3^2+ \sigma_1^3 \sigma_3^3+ \sigma_1^2 \sigma_2^3\sigma_4 )+ C \sigma_1\sigma_2 \sigma_3(\sigma_3^2+ \sigma_1^2 \sigma_4 )  \nonumber\\
  &+& D \sigma_2^4\sigma_4 + \sigma_1^4\sigma_4^2 + \sigma_3^4 + E \sigma_1\sigma_2^2 \sigma_3\sigma_4 +F\sigma_1^2\sigma_3^2\sigma_4+ G \sigma_2\sigma_4(\sigma_3^2+\sigma_1^2\sigma_4)\nonumber\\
  &+& H \sigma_2^2 \sigma_4^2 + I \sigma_1\sigma_3\sigma_4^2+ J \sigma_4^3,\label{k4}
\end{eqnarray}
where the constants are given in appendix A.

We have also computed the most  general polynomial
$Q_6(x_1,\ldots, x_6)$ which solves (\ref{qk}) with $k=4$. The
solution is an order 30 polynomial on the variables
$x_1,\ldots,x_6$ and too cumbersome to be reported here. For
$\Omega_4=0$ (we will see below why this choice is sensible),
$Q_6(x_1,\ldots, x_6)$ depends once more on a free parameter
$\Omega_6$, which as above acts as coefficient to the function
$K_6(x_1,\ldots,x_6)$ which satisfies the same equation
(\ref{kernel}) above.

Therefore, a structure seems  to emerge where the most general
$2k$-particle form factor depends on $k$ free parameters. A
similar structure was found when studying the boundary form
factors of specific fields in the $A_2$-affine Toda field theory
\cite{oota,CastroAlvaredo:2007pe}, although no physical
interpretation for the result was provided there. A more thorough
analysis of solutions to equations of the form (\ref{kernel}) was
carried out in \cite{Delfino:2006te} for the case $\alpha=-1$ and
the field $T\bar{T}$.

\subsection{Identifying the twist field form factors}
In this section we would like to argue that choosing  $\Omega_4=0$
in (\ref{q4}) corresponds to the specific twist field we are
interested in. The general solution (\ref{q4}) is a one-parameter
family of solutions characterized by the choice of the constant
$\Omega_4$. Given the usual assumption that the space of fields in
a local QFT is linear, we expect that the form factor of a linear
combination of fields is a linear combination of form factors,
that is, in general
\begin{equation}\label{fq}
    F^{\mathcal{O}_1+ \Omega \mathcal{O}_2}_k(x_1,\ldots,x_k)=
F^{\mathcal{O}_1}_k(x_1,\ldots,x_k)+ \Omega F^{\mathcal{O}_2}_k(x_1,\ldots,x_k)
\end{equation}
and therefore the solution (\ref{q4}) must describe the  form
factors of a linear combination of local fields (as would the
solution (\ref{sol2})). Since we are interested only in one very
particular field, the twist field $\mathcal{T}$, we must find a
suitable mechanism that allows us to select the particular value
of $\Omega_4$ corresponding to the four-particle form factor of
the twist field.

An interesting way of identifying the form factors of the twist
field is to use the form factor cluster decomposition
property, which has been studied for various models in the past
\cite{smirnov,Zamolodchikov:1990bk,KK,CastroAlvaredo:2000nk}
and analysed from a more general point of view in \cite{DSC}.
It is a factorization property of form factors which, for the
four particle case, can be expressed as
\begin{equation}
  \lim_{\kappa \rightarrow \infty}  F^{\mathcal{T}}_4(\kappa x_1,\kappa
  x_2,x_3,x_4)\propto F^{\mathcal{T}_1}_2( x_1,
  x_2)F^{\mathcal{T}_2}_2(x_3,x_4). \label{fact}
\end{equation}
In general, the fields $\mathcal{T}_1$ and $\mathcal{T}_2$ on the
r.h.s. may not necessarily correspond to the same field as the
form factor on the l.h.s. A notable example of this is the model
studied in \cite{CastroAlvaredo:2000nk} and the form factors of
the field $T\bar{T}$ studied in \cite{Delfino:2006te}. In
\cite{DSC} it was argued that for theories without internal
symmetries, the cluster decomposition would be a consequence of
the decoupling of right- and left-moving modes in the conformal
limit and would hold for any field whose counterpart in the
underlying conformal field theory is a primary field.

Given that the twist field does certainly correspond to a primary field in the underlying
conformal field theory we expect a factorization of the type (\ref{fact}).
Imposing (\ref{fact}) is in fact sufficient to select a single
value of $\Omega_4$ in (\ref{q4}). Indeed, if we carry out the
cluster limit in (\ref{fact}) for the general expression
(\ref{q4}) and we call $\sigma_i=\sigma_i(x_1,x_2)$ and
$\hat{\sigma}_i=\hat{\sigma}_i(x_3,x_4)$ we find that
\begin{eqnarray}
  \lim_{\kappa \rightarrow \infty} F_4(\kappa x_1,\kappa
  x_2,x_3,x_4) &\sim&  \left[\sigma_2 \hat{\sigma}_2 +
  \Omega_4(A \sigma_1^2 \hat{\sigma}_1^2 + B (\sigma_2\hat{\sigma}_1^2
  +\sigma_1^2 \hat{\sigma}_2) +
  D\sigma_2\hat{\sigma}_2)\right]\nonumber\\
  && \times
  \frac{F_{\text{min}}^{\mathcal{T}|11}(\frac{x_1}{x_2})F_{\text{min}}^{\mathcal{T}|11}(\frac{x_3}{x_4})
  }{(x_1-\alpha x_2)(x_2-\alpha x_1)(x_3-\alpha x_4)(x_4-\alpha x_3)}.
\end{eqnarray}
Clearly, this expression factorises if and only if $\Omega_4=0$.
In that case, we recover exactly (\ref{fact}) with
$\mathcal{T}_1=\mathcal{T}_2=\mathcal{T}$. We will therefore
choose $\Omega_4=0$ as our twist field solution.

If we had chosen to use the cluster decomposition property to fix
the constant $\Omega_2$ in (\ref{sol2}) we would have found
\begin{equation}
  \lim_{\kappa \rightarrow \infty}  F^{\mathcal{T}}_2(\kappa x_1,
  x_2)\propto \Omega_2, \label{fact2}
\end{equation}
so that our choice $\Omega_2=0$ guarantees that $\lim_{\kappa
\rightarrow \infty}   F^{\mathcal{T}}_2(\kappa x_1, x_2) \propto
F^{\mathcal{T}}_1 F^{\mathcal{T}}_1 =0$.

In general, it appears from our two-, four- and six-particle form
factor solutions that for every $a \in \mathbb{Z}^+$ there exists
a field $\mathcal{K}_{2a}$ whose form factors solve
(\ref{1})-(\ref{kre}) and have the interesting property that
\begin{equation}\label{zerof}
     F_0^{\mathcal{K}_{2a}}= F_2^{\mathcal{K}_{2a}|\mu_1\mu_2}
     (\theta_1,\theta_2)=\cdots=F_{2a-2}^{\mathcal{K}_{2a}|\mu_1\ldots\mu_{2a-2}}(\theta_1,\ldots,\theta_{2a-2})=0,
\end{equation}
consequently
$F_{2a}^{\mathcal{K}_{2a}|\mu_1\ldots\mu_{2a}}(\theta_1,\ldots,\theta_{2a})$
solves (\ref{kernel}) for $k=2a$, that is, it has no kinematic
poles.

At this stage we can only speculate about the nature of the fields
$\mathcal{K}_{2a}$. From the $a=2$ example, the cluster
decomposition property suggests that the field $\mathcal{K}_4$
does not correspond to a primary at conformal level, since at
least its four particle form factor does not factorize under
clustering. Furthermore, given that the form factors of all fields
$\mathcal{K}_{2a}$ are solutions to the twist field form factor
equations they must be
  twist fields of some kind. Finally, their non-vanishing form factors involve
  even particle numbers, which points to a particular kind of symmetry.
  One may think of linear combinations of composite fields such as
  $\Theta \mathcal{T}$ or $\varphi^2
\mathcal{T}$ etc. As a future project it would be very interesting
to identify the precise nature of the fields $\mathcal{K}_{2a}$.

\section{Twist field form factors for the $SU(3)_2$-HSG model}
We turn now to the second model whose twist field form factors we
wish to investigate. The complexity of the model is increased by a
number of features, notably the fact that its spectrum has two
particles and the presence of the free parameter $\sigma$ in the
$S$-matrix (\ref{hsg}). The solutions to the equations
(\ref{fmin}) with the S-matrices
(\ref{hsg}) are
\begin{equation}
    F_{\text{min}}^{\pm \pm}(\theta)=-i \sinh\left(\frac{\theta}{2n}\right),
\label{ising}
\end{equation}
and
\begin{equation}\label{min2}
    F_{\text{min}}^{\pm \mp}(\theta)=A(n) e^{\pm \frac{\theta}{4n}+ \frac{i\pi(1 \mp 1)}{4}} \exp\left(\int_{-\infty}^\infty \frac{dt}{t}\frac{\sinh^2\left(\frac{t}{2}\left(n+ \frac{i (\theta \pm \sigma)}{\pi}\right) \right)}{\sinh(nt)\cosh(t/2)}\right),
\end{equation}
with $A(n)$ given by the limit,
\begin{equation}
A(n)=\lim_{p \rightarrow \infty} e^{-\frac{2p+2+n}{2n}- \frac{i\pi}{4}}\sqrt{\frac{2n}{p}}\left(\frac{4p+3+2n}{4n} \right)^{\frac{4p+3+2n}{8n} }\left(\frac{4p+5+2n}{4n} \right)^{\frac{4p+5+2n}{8n} } \prod_{k=0}^p \frac{\Gamma\left(\frac{4k+1 +2n}{4n}\right)^2}{\Gamma\left(\frac{4k+3+2n}{4n}\right)^2}.
\end{equation}
The solution (\ref{ising})  is nothing but the Ising model
solution first obtained in \cite{entropy}, as we would expect from
the first S-matrix in (\ref{hsg}).

The form factors (\ref{min2}) can also be expressed in terms of an
infinite product of Gamma functions
\begin{equation}\label{ming}
    F_{\text{min}}^{\pm \mp}(\theta)=A(n) e^{\pm \frac{\theta}{4n}+ \frac{i\pi(1 \mp 1)}{4}} \prod_{k=0}^\infty \frac{\Gamma\left(\frac{4k+3+2n}{4n}\right)^2\Gamma\left(\frac{-2 w+4k+1+2n}{4n}\right)\Gamma\left(\frac{2 w+4k+1+2n}{4n}\right)}{\Gamma\left(\frac{4k+1+2n}{4n}\right)^2\Gamma\left(\frac{-2 w+4k+3+2n}{4n}\right)\Gamma\left(\frac{2 w+4k+3+2n}{4n}\right)},
\end{equation}
with $w=n+i (\theta\pm \sigma)/{\pi}$.

The function $A(n)$ defined above would seem a strange choice of normalization. The motivation for it is to  ensure that the following minimal form factor relations
\begin{equation}\label{re}
     F_{\text{min}}^{\pm \mp}(\theta) F_{\text{min}}^{\pm \mp}(\theta+ i \pi)=
     \pm \frac{e^{\pm \frac{\theta}{2n}\pm \frac{i \pi}{4n}}}{\sinh\left(\frac{\theta\pm\sigma}{2n}+\frac{i\pi}{4n} \right)},
\end{equation}
hold, without involving complicated constants. In particular, $A(1)=e^{-G/\pi} e^{-i\pi/4} 2^{1/4}$, where $G$ is the Catalan constant that appears in the normalization of the form factors of the one-copy model \cite{Delfino:1994ea,CastroAlvaredo:2000em}. It is worth noticing however that with respect to the latter normalization our minimal form factor at $n=1$ is multiplied by the extra factor $e^{-i \pi/4}$.

Once the two-particle form factor and minimal form factor have
been computed the basic monodromy and pole structure features of
the form factors are fixed so that higher particle form factors
can be constructed in terms of the solutions already found. Let us
introduce the following notation:
\begin{equation}
   F_{\ell+m}(\{x\}_\ell^{+};\{x\}_m^{-}):=
   F_{\ell+m}^{\mathcal{T}|\overbrace{+ \ldots +}^\ell \overbrace{- \ldots -}^m}
 (x_1,\ldots,x_\ell,x_{\ell+1}\ldots x_{\ell+m}),
\end{equation}
 This represents the $\ell+m$-particle
form factor of the twist field with $\ell$ particles of type $+$
and $m$ particles of type $-$ living in one particular copy of the
model. For the model under consideration, we will make the
following ansatz
\begin{eqnarray}
  F_{\ell+m}(\{x\}_\ell^{+};\{x\}_m^{-})&=&H_{\ell,m}^{+-} Q_{\ell+m}^{+-}(\{x\}_\ell^{+};\{x\}_m^{-})
   \prod_{1\leq i<j\leq \ell}
\frac{F_{\text{min}}^{\mathcal{T}|++}(\frac{x_i}{x_j})}{(x_i-\alpha
x_j)(x_j-\alpha x_i)}\nonumber \\
&& \times \prod_{i=1}^\ell
\prod_{j=\ell+1}^{\ell+m}F_{\text{min}}^{\mathcal{T}|+-}(\frac{x_i}{x_j})
\prod_{\ell+1\leq i<j\leq \ell+m}
\frac{F_{\text{min}}^{\mathcal{T}|--}(\frac{x_i}{x_j})}{(x_i-\alpha
x_j)(x_j-\alpha x_i)}.\label{ansatz}
\end{eqnarray}
In terms of the new variables $x_i$
we can rewrite for example
\begin{equation}
   F_{\text{min}}^{\mathcal{T}|\pm\pm}(\theta_i-\theta_j)\equiv
   F_{\text{min}}^{\mathcal{T}|\pm\pm}(\frac{x_i}{x_j}).
\end{equation}
It is easy to check that, the ansatz (\ref{ansatz}) automatically
satisfies equations (\ref{1}) and (\ref{2}) provided that the
functions $Q_{\ell+m}^{+-}(\{x\}_\ell^{+};\{x\}_m^{-})$ are
separately symmetric in both sets of variables and have no poles
on the physical sheet and $H_{\ell,m}^{+-}$ are rapidity
independent. Notice that there are kinematic poles associated to
pairs of $+$ and $-$ particles, but not to the combination $+-$,
as the two particles in the model are self-conjugated (their own
antiparticle). The ansatz (\ref{ansatz}) is reminiscent of the
solution procedure used in
\cite{CastroAlvaredo:2000em,CastroAlvaredo:2000nk}  where the form
factors of local fields of the  present model were also studied.

Once the ansatz (\ref{ansatz}) has been made it remains to
identify the functions
$Q_{\ell+m}^{+-}(\{x\}_\ell^{+};\{x\}_m^{-})$ and the constants
$H_{\ell,m}^{+-}$. A useful benchmark that can be employed for
this model is the fact that whenever $m=0$ or $\ell=0$, the
resulting form factor must be the $\ell$-particle or $m$-particle
form factor of the Ising model, respectively. This relationship
with the Ising model, combined with the kinematic residue equation
(\ref{3}) also implies that only form factors with both $\ell$ and
$m$ even will be non-vanishing.

Substituting the ansatz (\ref{ansatz}) into (\ref{3}) we obtain
the following recursive relations for
$Q_{\ell+m}^{+-}(\{x\}_\ell^{+};\{x\}_m^{-})$ and the constants
$H_{\ell,m}^{+-}$,
\begin{equation}
    H_{\ell+2,m}^{+-}=\frac{
    \alpha^{\frac{3\ell-m}{2}+2}e^{-\frac{\sigma m}{2n}}2^{2\ell-m+1}
    \sin\frac{\pi}{n}}{n
    F_{\text{min}}^{\mathcal{T}|++}(i\pi)}\,
    H_{\ell,m}^{+-},\label{h1}
\end{equation}
and
\begin{equation}
   Q_{\ell+2+m}^{+-}(\alpha x_0,
   x_0,\{x\}_\ell^{+};\{x\}_m^{-})=P_{\ell,m}^{+-}(x_0,\{x\}_\ell^{+};\{x\}_m^{-})
   Q_{\ell+m}^{+-}(\{x\}_\ell^{+};\{x\}_m^{-}),\label{q3}
\end{equation}
with
\begin{equation}
    P_{\ell,m}^{+-}(x_0,\{x\}_\ell^{+};\{x\}_m^{-})=\alpha^\ell x_0^{\ell+2-m}\sigma_\ell^+
   \sum_{i,j=0}^\ell\left(-\frac{x_0}{\alpha}\right)^{\ell-i}(-\alpha^2
   x_0)^{\ell-j}\sigma_i^+ \sigma_j^+ \sum_{k=0}^{m}(-\sqrt{\alpha}
   e^{\frac{\sigma}{n}}x_0)^{m-k}\sigma_k^-,\label{q4hsg}
\end{equation}
where $\sigma_k^{+}, \sigma_k^{-}$ are elementary symmetric
polynomials on the variables $\{x\}_\ell^{+}$ and $\{x\}_m^{-}$,
respectively. To simplify notation, in (\ref{q4hsg}) and
(\ref{q2}) we have dropped the explicit variable dependence of the
symmetric polynomials.

In the ansatz (\ref{ansatz}) we have chosen a particular ordering
of the particles with type + appearing first and type - last. Of
course this ordering can be changed by employing the first form
factor equation (\ref{1}). Alternatively, we could have worked
with the form factor $F_{\ell+m}(\{x\}_\ell^{-};\{x\}_m^{+})$
where we now have $\ell$ particles of type - first, followed by
$m$ particles of type +. For this ordering, the recurrence
equations above become instead
\begin{equation}
    H_{\ell+2,m}^{-+}=\frac{
    \alpha^{\frac{3\ell}{2}+2}e^{-\frac{\sigma m}{2n}}2^{2\ell-m+1}
    \sin\frac{\pi}{n}}{n
    F_{\text{min}}^{\mathcal{T}|--}(i\pi)}\,
    H_{\ell,m}^{-+},\label{h2}
\end{equation}
and
\begin{equation}
   Q_{\ell+2+m}^{-+}(\alpha x_0,
   x_0,\{x\}_\ell^{-};\{x\}_m^{+})=P_{\ell,m}^{-+}(x_0,\{x\}_\ell^{-};\{x\}_m^{+})
   Q_{\ell+m}^{-+}(\{x\}_\ell^{-};\{x\}_m^{+}),\label{q1}
\end{equation}
with
\begin{equation}
    P_{\ell,m}^{-+}(x_0,\{x\}_\ell^{-};\{x\}_m^{+})=\alpha^\ell
    x_0^{\ell+2}\frac{\sigma_\ell^-}{\sigma_m^+}
   \sum_{i,j=0}^\ell\left(-\frac{x_0}{\alpha}\right)^{\ell-i}(-\alpha^2
   x_0)^{\ell-j}\sigma_i^- \sigma_j^- \sum_{k=0}^{m}(-\sqrt{\alpha}
   e^{\frac{\sigma}{n}}x_0)^{m-k}\sigma_k^+.\label{q2}
\end{equation}
From the definition (\ref{ansatz}) and equations (\ref{1}) and (\ref{fmin}) it is easy to show that
\begin{equation}\label{equal}
    H_{\ell,m}^{+-}Q_{\ell,m}^{+-}(\{x\}_\ell;\{x\}_m)=
    H_{m,\ell}^{-+}Q_{m,\ell}^{-+}(\{x\}_m;\{x\}_\ell),
\end{equation}
which provides a useful relationship between the solutions of (\ref{q1}) and those of (\ref{q3}).

\subsection{Solutions to the recursive equations}
Given the structure of the $S$-matrix (\ref{hsg}) we know that
form factors involving only particles of type + or only particles
of type - should equal the form factors of the Ising model. We
will therefore split our solutions into Ising model solutions and
solutions involving particles of both types.
\subsubsection{Ising model solutions}
For $m=0$ in (\ref{h1})-(\ref{q3}) or equivalently $m=0$ in
(\ref{h2})-(\ref{q1}) the equations reduce to the form factor
equations of the Ising model. That is,
\begin{equation}
    H_{\ell+2}=\frac{
    \alpha^{\frac{3\ell}{2}+2}2^{2\ell+1}
    \sin\frac{\pi}{n}}{n
    F_{\text{min}}^{\mathcal{T}|\pm \pm}(i\pi)}\,
    H_{\ell}, \label{h2ising}
\end{equation}
and
\begin{equation}
   Q_{\ell+2}(\alpha x_0,
   x_0,\{x\}_\ell)=P_{\ell}(x_0,\{x\}_\ell)
   Q_{\ell}(\{x\}_\ell),\label{q1ising}
\end{equation}
with
\begin{equation}
    P_{\ell}(x_0,\{x\}_\ell)=\alpha^\ell
    x_0^{\ell+2}{\sigma_\ell}
   \sum_{i,j=0}^\ell\left(-\frac{x_0}{\alpha}\right)^{\ell-i}(-\alpha^2
   x_0)^{\ell-j}\sigma_i \sigma_j.\label{q2ising}
\end{equation}
Interestingly even for the Ising model, these equations are not
easy to solve and the solutions for $Q_{\ell}(\{x\}_\ell)$ become
very cumbersome beyond $\ell=4$. The first few solutions are,
\begin{eqnarray}
Q_2(x_1,x_2)&=& \alpha^{-1} \sigma_2,\label{2par}\\
Q_{4}(x_1,x_2,x_3,x_4) &=&\alpha^{-1}\sigma_4^2\left(\sigma_2^2
-\frac{p_1(\alpha)\sigma_1\sigma_3 }{\alpha}+\frac{p_1(\alpha)
(1+\alpha^{2})^2\sigma_4}{\alpha^3}\right),\label{fr1}\\
Q_6(x_1,x_2,x_3,x_4,x_5,x_6)&=& {\sigma }_6^3\left({\sigma
}_2^2{\sigma }_4^2 +
    \frac{ p_1(\alpha)^2{{\sigma }_1}{\sigma }_3^2
       {{\sigma }_5}}{{\alpha }^2} + \frac{p_2(\alpha)
       {{\sigma }_1}{{\sigma }_2}{{\sigma }_4}{{\sigma }_5}}{\alpha }\right.\nonumber\\
        && \left.-
    \frac{{\left( 1 + {\alpha }^2 \right) }^2
       p_3(\alpha)  {\sigma }_1^2
       {\sigma }_5^2}{{\alpha }^4}-
    \frac{ p_1(\alpha) {{\sigma }_3}
       \left( {{\sigma }_1}{\sigma }_4^2 + {\sigma }_2^2{{\sigma }_5} \right) }{\alpha
       } \right.\nonumber\\
        && \left. + \frac{ p_2(\alpha)
       {p_1(\alpha)}^4{\sigma }_3^2{{\sigma }_6}}{{\alpha }^
       5}  + \frac{{\left( 1 + {\alpha }^2 \right) }^2
       p_1(\alpha)
       \left( {\sigma }_4^3 + {\sigma }_2^3{{\sigma }_6} \right) }
       {{\alpha }^3}\right.\nonumber\\
        && \left.-
    \frac{ p_2(\alpha)
       { p_1(\alpha)  }^2
       p_4(\alpha)
       p_3(\alpha)  {{\sigma }_1}
       {{\sigma }_5}{{\sigma }_6}}{{\alpha }^7} +
    \frac{{ p_2(\alpha)}^3
       {p_1(\alpha)}^4
       p_3(\alpha) {\sigma }_6^2}
       {{\alpha }^9} \right.\nonumber\\
        && \left.- \frac{{ p_1(\alpha) }^2
      p_3(\alpha) {{\sigma }_3}
       \left( {{\sigma }_4}{{\sigma }_5} + {{\sigma }_1}{{\sigma }_2}{{\sigma }_6} \right) }
       {{\alpha }^4}- \frac{ p_2(\alpha)
       {p_1(\alpha) }^2
       p_5(\alpha)
       {{\sigma }_2}{{\sigma }_4}{{\sigma }_6}}{{\alpha }^5}\right.\nonumber\\
        && \left.  +
    \frac{p_1(\alpha)
       {p_3(\alpha)}^2
       \left( {{\sigma }_2}{\sigma }_5^2 +
         {\sigma }_1^2{{\sigma }_4}{{\sigma }_6} \right) }{{\alpha }^5}
         \right), \label{q6}
\end{eqnarray}
with
\begin{eqnarray}
  p_1(\alpha) &=& 1 + \alpha  + {\alpha }^2, \nonumber \\
  p_2(\alpha) &=& 1 - \alpha  + {\alpha }^2 \nonumber \\
  p_3(\alpha) &=& 1 + \alpha  + {\alpha }^2 + {\alpha }^3 + {\alpha }^4,\nonumber \\
  p_4(\alpha) &=& 1 - \alpha  + 3{\alpha }^2 - {\alpha }^3 + {\alpha }^4
  ,\nonumber \\
  p_5(\alpha) &=& 3 + 2\alpha  + 4{\alpha }^2 + 2{\alpha }^3 + 3{\alpha
  }^4.
\end{eqnarray}
However the twist field form factors are already know for the
Ising model. They were computed in \cite{nexttonext} not by
solving equations (\ref{1})-(\ref{kre}) but by using the special
free fermion features of the model. It was found that
\begin{equation}\label{fising}
    F_{\ell}^{\mathcal{T}}(\theta_1,\ldots,\theta_\ell)=\text{Pf}(K^{(\ell)}),
\end{equation}
where $\text{Pf}(K^{(\ell)})=\sqrt{\det(K^{(\ell)})}$ and
$K^{(\ell)}$ is an $\ell \times \ell$ matrix whose entries are
given by $
K^{(\ell)}_{ij}=F_2^{\mathcal{T}|\pm\pm}(\theta_{ij})/{\langle\mathcal{T}\rangle}.
$ Comparing to our original ansatz we have the remarkable identity
\begin{equation}
    Q_{\ell}(\{x\}_{\ell})=H_{\ell}^{-1} \text{Pf}(K^{\ell})\prod_{i<j}^{\ell}(x_i-\alpha
    x_j)(x_j-\alpha x_i). \label{rel}
\end{equation}
Bringing the r.h.s. of (\ref{rel}) into the form of a combination
of symmetric polynomials is highly non-trivial for $\ell>4$. In
particular, for $\ell=6$ it yields the result (\ref{q6}). It would
be nice to develop a general proof of (\ref{rel}).

\subsection{Solutions involving both particle types}
Starting with the two particle solutions (\ref{2par}) we find the
following new four particle form factor solutions
\begin{eqnarray}
  Q_{2+2}^{+-}(x_1,x_2;x_3,x_4) &=& \alpha^{-1}\sigma_2^-\left(\hat{\sigma}_2^+
  -\frac{\sqrt{\alpha}}{1+\alpha} \hat{\sigma}_1^+\sigma_1^- + \sigma_2^- \right), \\
   Q_{2+2}^{-+}(x_1,x_2;x_3,x_4) &=& \alpha^{-2}\sigma_2^+
   \left(\hat{\sigma}_2^-
  -\frac{\sqrt{\alpha}}{1+\alpha} \sigma_1^+\hat{\sigma}_1^- + \sigma_2^+
  \right),\label{fourpar}
\end{eqnarray}
where $\hat{\sigma}_k^{\pm}$ are symmetric polynomials on the
variables $\{x e^{\frac{\sigma}{n}}\}_{\ell,m}$.

Going beyond four particles is rather difficult, but because of
their relationship to the form factors of the Ising model, it is
possible to find closed formulae for certain types of form
factors. For example, when $\ell=2$ and $m$ is general. In this
particular case the form factor equations become simply
\begin{equation}
   Q_{2+m}^{+-}(\alpha x_0,
   x_0;\{x\}_m^{-})=x_0^{2-m}
   \sum_{k=0}^{m}(-\sqrt{\alpha}
   e^{\frac{\sigma}{n}}x_0)^{m-k}\sigma_k^-
   Q_{m}(\{x\}_m^{-}),\label{q32}
\end{equation}
and
\begin{equation}
   Q_{2+m}^{-+}(\alpha x_0,
   x_0;\{x\}_m^{+})=\frac{x_0^{2}}{\sigma_m^+}
   \sum_{k=0}^{m}(-\sqrt{\alpha}
   e^{\frac{\sigma}{n}}x_0)^{m-k}\sigma_k^+
   Q_{m}(\{x\}_m^{+}),\label{q33}
\end{equation}
where $Q_{m}(\{x\}_m)$ is the Ising model solution given by
(\ref{rel}). Particular solutions to (\ref{q32}) and (\ref{q33})
take the form,
\begin{equation}
   Q_{2+m}^{+-}(x_1,
   x_2;\{x\}_m^{-})=\left[\alpha^{\frac{m}{2}-1}\hat{ \sigma}_2^+
   \sum_{k=0}^{\frac{m}{2}}\frac{{\sigma}_{2k}^-}{(\hat{\sigma}_2^+)^{k}}
   -\frac{\alpha^{\frac{m-1}{2}}\hat{\sigma}_1^+}{1+\alpha}\sum_{k=0}^{\frac{m-2}{2}}
   \frac{\sigma_{2k+1}^-}{(\hat{\sigma}_2^+)^{k}}\right]
   Q_{m}(\{x\}_m^{-}),\label{sol1}
\end{equation}
and
\begin{equation}
   Q_{2+m}^{-+}(x_1,
   x_2;\{x\}_m^{+})=\frac{(\hat{\sigma}_2^-)^{\frac{m}{2}}}{\sigma_m^+}\left[\alpha^{-1}
  \hat{ \sigma}_2^-
   \sum_{k=0}^{\frac{m}{2}} \frac{\hat{\sigma}_{2k}^+}{(\hat{\sigma}_2^-)^{k}}
   -\frac{\alpha^{-\frac{1}{2}}\hat{\sigma}_1^-}{1+\alpha} \sum_{k=0}^{\frac{m-2}{2}}
  \frac{\sigma_{2k+1}^+}{(\sigma_2^-)^{k}}\right]
   Q_{m}(\{x\}_m^{+}).\label{sol2hsg}
\end{equation}
They provide closed solutions to the equations
(\ref{q32})-(\ref{q33}) valid for any values of $m$.
Unfortunately, this is not enough to conclude they are fully
consistent with all form factor equations. What we mean is that
the relation (\ref{equal}) must also hold, which means that for
example the solution $Q_{2+4}^{-+}(x_1,
   x_2;\{x\}_4^{+})$ constructed above, must also solve the form
   factor equation satisfied by
   $Q_{4+2}^{+-}(\{x\}_4^{+};x_1,x_2)$ (up to constants). This imposes a set of
   further constraints on the solutions to (\ref{q32}) and
   (\ref{q33}). Let us consider an example.

From (\ref{sol1}) we find
\begin{equation}\label{Q24}
 Q_{2+4}^{-+}(x_1,
   x_2;\{x\}_4^{-})=\alpha^{-1} \frac{\hat{\sigma}_2^-}{\sigma_4^+}\left[
   (\hat{ \sigma}_2^-)^2 +{{\hat{\sigma}}_{2}^-}{{\sigma}_{2}^+}+
   {{\sigma}_{4}^+}
   -\frac{\sqrt{\alpha}\hat{\sigma}_1^-}{1+\alpha}\left(\hat{\sigma}_{2}^- \sigma_{1}^+ +
  {\sigma_{3}^+}\right)\right]
   Q_{4}(\{x\}_4^{+}).
\end{equation}
This function solves (\ref{q32}), however it does not solve the
equation for $Q_{4+2}^{+-}(\{x\}_4^{+};x_1,x_2)$ which can be
obtained from (\ref{q2}) with $\ell=2$, $m=4$. If we solve that
equation, we obtain a completely different solution. Therefore
(\ref{Q24}) is not a consistent solution to all form factor
equations.  As we studied in detail for the RT-model, we can generally
add an extra function to any solution, as long as that function is
in the kernel of the equation we are trying to solve. In our case,
this means that we can always add to (\ref{Q24}) any function
$K_{2+4}^{-+}(x_1,x_2;\{x\}_4)$ which satisfies,
\begin{equation}
   K_{2+4}^{-+}(\alpha x_0,
   x_0;\{x\}_4^{+})=0.
\end{equation}
The most general solution to this equation, up to a multiplicative
constant is,
\begin{eqnarray}
&&  K_{2+4}^{-+}(\alpha x_0,
   x_0;\{x\}_4^{+})=\frac{\sigma_2^- \sigma_4^+ (\alpha (\sigma_1^+)^2 -(\alpha+1)^2 \sigma_2^+)}{\alpha^3(1+\alpha)^2}\nonumber\\
 && \times \left(\alpha^3 \sigma_1^+\sigma_2^+ \sigma_3^+ -\alpha(1+\alpha+\alpha^2)((\sigma_3^+)^2+(\sigma_1^+)^2\sigma_4^+) + (1+\alpha)^4(1+\alpha^2)\sigma_2^+\sigma_4^+) \right).
\end{eqnarray}
Solving for $Q_{4+2}^{+-}(\{x\}_4^{+};x_1,x_2)$ we find that
\begin{eqnarray}\label{Q42}
Q_{4+2}^{+-}(\{x\}_4^{+};x_1,x_2)&=& \alpha
\frac{\hat{\sigma}_2^-}{\sigma_4^+}\left[
   (\hat{ \sigma}_2^-)^2 +{{\hat{\sigma}}_{2}^-}{{\sigma}_{2}^+}+
   {{\sigma}_{4}^+}
   -\frac{\sqrt{\alpha}\hat{\sigma}_1^-}{1+\alpha}\left(\hat{\sigma}_{2}^- \sigma_{1}^+ +
  {\sigma_{3}^+}\right)\right]
   Q_{4}(\{x\}_4^{+})\nonumber\\
   &&+K_{2+4}^{-+}(x_1,
   x_2;\{x\}_4^{+}),
\end{eqnarray}
and from equation (\ref{equal}) it follows that \beq
Q_{2+4}^{-+}(x_1,
   x_2;\{x\}_4^{-})=\alpha^{-2}Q_{4+2}^{+-}(\{x\}_4^{+};x_1,x_2).
\eeq
 Therefore, in general, the solutions (\ref{sol1}) and
(\ref{sol2hsg}) need to be modified by adding some function in the
kernel of (\ref{q32}) or (\ref{q33}) which is consistent with
(\ref{equal}).
\section{Numerical results \label{sec:numerical results}}
In this section we want to provide numerical evidence that the
twist field form factors computed thus far do indeed correspond to
the correct twist field. Our method is to  check numerically the
form factors of the two models presented in sections \ref{rtmodel}
and \ref{sgmodel} against the $\Delta$-sum rule.
For both models, we considered only the
two and four particle terms in the expansion (\ref{exexpansion1}).

For each model, we have adopted a different numerical approach: For the RT-model we have only evaluated (\ref{delta2}) directly in the ultraviolet limit $r_0=0$. This is because the numerical recipe used  was particularly slow. We employed a truncated version of the infinite product of Gamma functions  given in \cite{entropy} to evaluate the minimal form factor (\ref{int}). We also employed a Newton-Cotes method and a Montecarlo simulation which uses the Vegas algorithm to perform the integrals in the two-particle and four-particle case, respectively.

For the HSG-model though we have been able to consider a wide range of values of $r_0$ by employing a very precise,
piece-wise polynomial interpolation of the functions
$F_{\text{min}}^{\mathcal{T}|\pm\mp}(\theta)F_{\text{min}}^{\Theta|\pm\mp}(\theta)^*
$ which dramatically reduces the running time of our programme.
In this case we have carried out
the integrals by means of the Vegas algorithm.

The numerical results are shown below.

\subsection{Numerical results for the roaming trajectories model}
As explained in section \ref{rtmodel}, the function $c_{\text{eff}}(r)$ or effective central charge of this model exhibits an infinite set of plateaux between $r=0$ and $r\rightarrow \infty$. A similar type of behaviour is expected for $\Delta^{\mathcal{T}}(r_0)$ as $r_0$ is varied. Here we have only considered $r_0=0$, however our results still allow us to identify two values of $\Delta^{\mathcal{T}}$, that is the value obtained in the two-particle approximation and the value obtained in the four-particle approximation. Each of these values agrees with what we would have expected for the first two plateaux of the function $\Delta^{\mathcal{T}}(r_0)$.

The precise location of the plateaux can be easily predicted by combining (\ref{cc}) with (\ref{known1}). The first two values of the central charge
correspond to $c_{3}=1/2$ and $c_{4}=7/10$, respectively. Inserting these values in (\ref{known1}) we obtain a value of $\Delta^{\mathcal{T}}$ for each central charge and each value of $n$.

The two particle contribution takes exactly the same form as for the sinh-Gordon model and was given in \cite{entropy}. Evaluating it for $\theta_0=20$ we obtain the values listed below
 \begin{table}[h]
 \label{2part}
 \begin{center}
 \begin{tabular}{| l | c | c | }
  \hline
  &&\\
   $n$ & $\frac{1}{48}\left(n-\frac{1}{n}\right)$ & $ \Delta_{2}^{\mathcal{T}}(0)$  \\
  &&\\
 \hline
 \hline 2 &  0.03125 & 0.0312548\\
 \hline 3 & 0.0555556 & 0.055676  \\
 \hline 4 & 0.078125 & 0.0785953 \\
 \hline 5 &  0.1 & 0.101033  \\
 \hline 6 & 0.121528 & 0.123257    \\
 \hline 7 & 0.142857 & 0.145351 \\
 \hline 8 &   0.164062 &0.167351\\
 \hline 9 & 0.185185 & 0.189277 \\
 \hline 10 & 0.20625 &  0.211143 \\
 \hline
 \end{tabular}
 \end{center}
 \caption{Two particle contribution to the conformal dimension in the RT-model. The second column shows the exact values of the conformal dimension of the twist field corresponding to central charge $c_3=1/2$. The third column shows the numerical values of the same quantity in the two-particle approximation for $\theta_{0}=20$.}
 \end{table}\\
Employing the four-particle form factors of the energy-momentum
tensor  obtained in \cite{FMS} and \cite{KK} and our solution
(\ref{q4}) with $\Omega_4=0$, the four-particle contribution is
given by
\begin{eqnarray}
\Delta^{\mathcal{T}}_{4}(0)&=&-\frac{\sin \left( \frac{\pi}{n} \right) \cosh \left( \theta_0 \right) \left| F_{\text{min}}^{\mathcal{T}}(i\pi)^* F_{\text{min}}^{\Theta}(i\pi) \right|^{2}}{1536\pi^{3}}\nonumber \\
&&
\int_{-\infty}^{\infty}d\theta_1 d\theta_2 d\theta_3 d \theta_4 \frac{\sigma_1 \sigma_2 \sigma_3 Q_{4}(x_{1},x_{2},x_{3},x_{4})\prod\limits_{i<j}F_{\text{min}}^{\mathcal{T}}(\theta_{ij})^{*}F_{\text{min}}^{\Theta}(\theta_{ij})}{ \left(\prod\limits_{i<j}  \cos \left( \frac{\pi}{n} \right)-\cosh \left( \frac{\theta_{ij}}{n} \right) \right)
(\sum\limits_{i=1}^{4}\cosh(\theta_{ij}))^{2}} \label{4prtm}
\end{eqnarray}
where $\sigma_1, \sigma_2$ and $\sigma_3$ above represent
 elementary symmetric polynomials in the variables $e^{\theta_i}$
 with $i=1,2,3,4$. The values of (\ref{4prtm}) for different values of
  $n$ are given in Table 2.
 \begin{table}[h]
 \label{4part}
 \begin{center}
 \begin{tabular}{| l | c | c | }
  \hline
  &&\\
   $n$ & $\frac{1}{120}\left(n-\frac{1}{n}\right)$ & $ \Delta_{4}^{\mathcal{T}}(0)$  \\
  &&\\
 \hline
 \hline 2 &  0.012500 & 0.013086 \\
 \hline 3 & 0.022200 & 0.022169 \\
 \hline 4 & 0.031250 & 0.028611\\
 \hline 5 &  0.040000 & 0.042555 \\
 \hline 6 & 0.048611 & 0.047566\\
 \hline 7 & 0.057143 & 0.057996 \\
 \hline 8 &   0.065625 & 0.064736 \\
 \hline 9 & 0.074074 & 0.072281\\
 \hline 10 & 0.082500 & 0.068762 \\
 \hline
 \end{tabular}
 \end{center}
 \caption{Four-particle contribution to the conformal dimension in the RT-model. The second column shows the difference between the values of the conformal dimension of the twist field corresponding to central charges $c_4=7/10$ and $c_2=1/2$. The third column shows the numerically computed four-particle contribution to the conformal dimension  for $\theta_{0}=20$.}
 \end{table}
Both Table 1 and 2 show relatively good agreement between the
values predicted by the theory and those numerically obtained. The
difference between the theoretical and numerical values is
considerable for some of the results in Table 2, specially as $n$
is increased. However it is always within the standard deviation
of the computation.
\subsection{$\Delta$-sum rule for the $SU(3)_{2}$-HSG model}
From (\ref{full}) and the two particle form factor of the
energy-momentum tensor for the thermally perturbed Ising model
 \begin{equation}
 F_{2}^{\Theta |\pm \pm}(\theta)=-2\pi
 im^{2}\sinh(\frac{\theta}{2}),
   \label{2enmom}
\end{equation}
the two particle contribution can be easily calculated to
\begin{equation}
\Delta^{\mathcal{T}}_{2}(\tilde{r}_{0}) =\frac{2\cos\left(
\frac{\pi}{2n}
\right)}{4\pi}\int_{-\infty}^{\infty}d\theta_{1}d\theta_{2}\frac{(1+\tilde{r}_{0}\sum_{i=1}^2\cosh(\theta_{i}))
e^{-\tilde{r}_{0}\sum\limits_{i=1}^2\cosh(\theta_{i})}}{(\sum_{i=1}^2\cosh(\theta_{i}))^{2}}\frac{\sinh
\left(   \frac{\theta_{12}}{2n}  \right)\sinh  \left(
\frac{\theta_{12}}{2}  \right)}{\cosh \left( \frac{\theta_{12}}{n}
\right) -\cos \left(  \frac{\pi}{n} \right)},
 \label{2contr}
\end{equation}
where $\tilde{r}_{0}=mr_{0}$ is a dimensionless parameter
proportional to the mass scale. From a physical point
of view, we expect this contribution to produce a function with a
plateau at exactly
$\Delta_2^{\mathcal{T}}(0)=\frac{1}{24}\left(n-\frac{1}{n}\right)$,
which is the value corresponding to two copies of the Ising model
or $c=1$.
\begin{figure}[h!]
\includegraphics[width=0.52\textwidth]{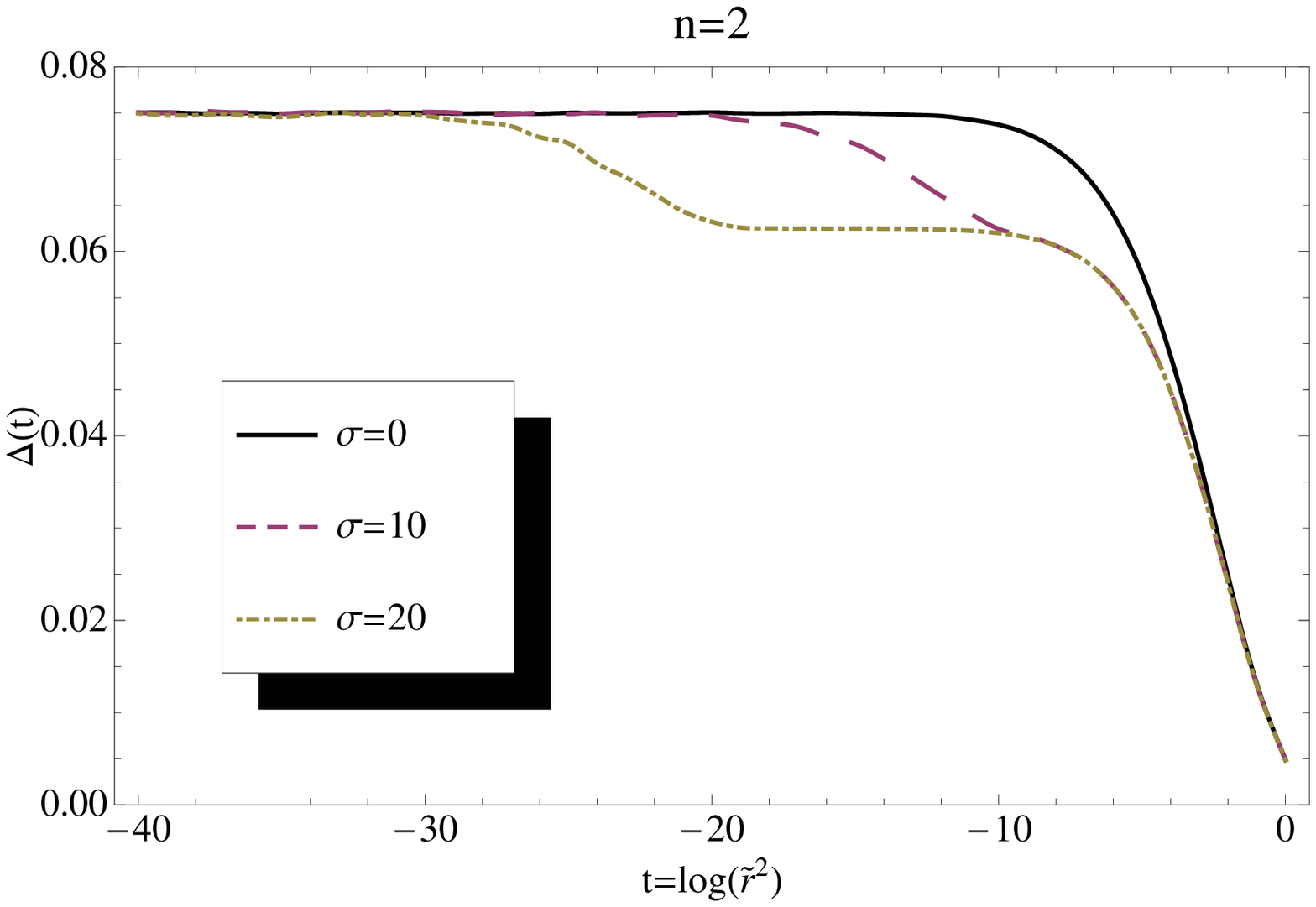}
\includegraphics[width=0.52\textwidth]{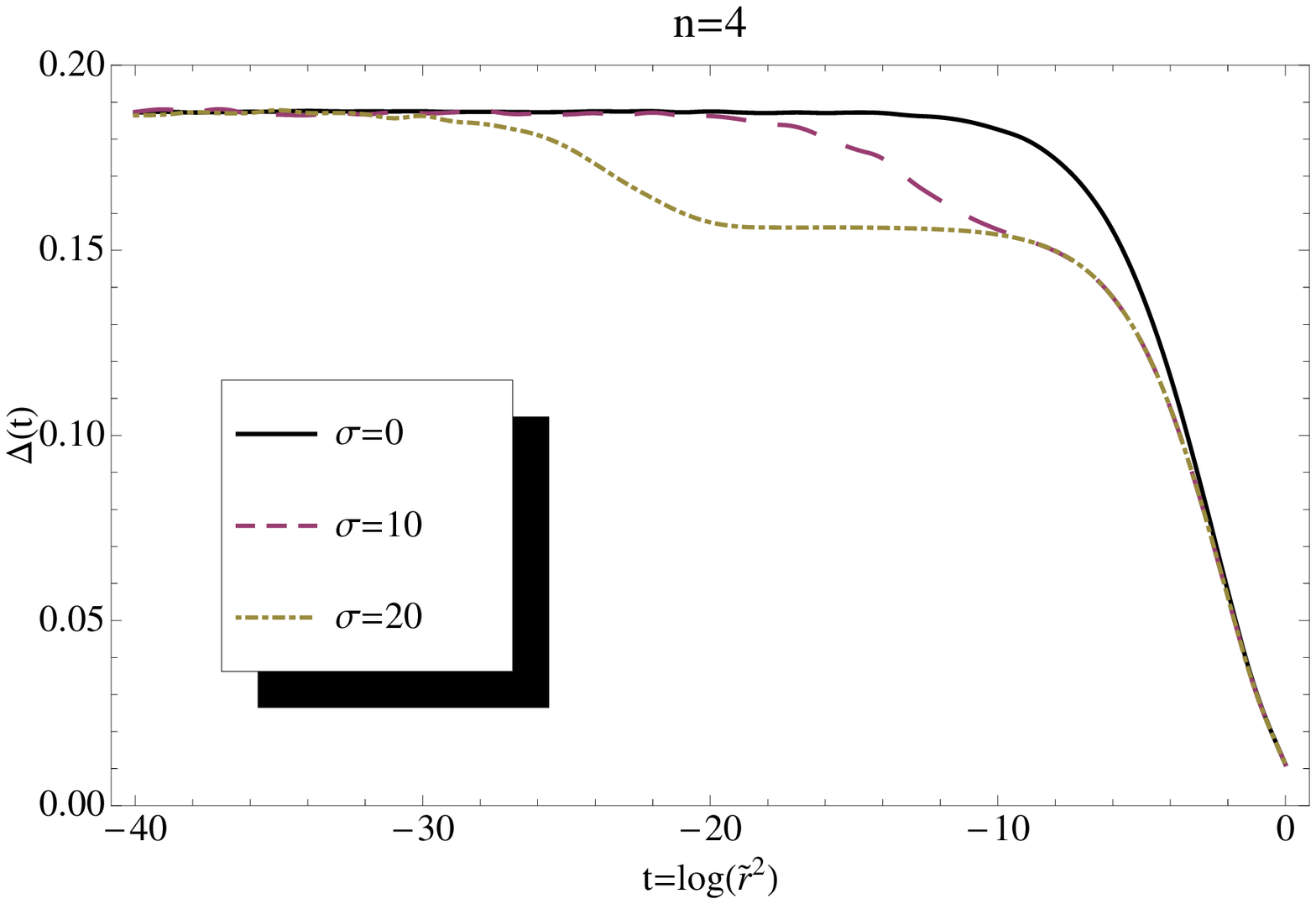}
\includegraphics[width=0.52\textwidth]{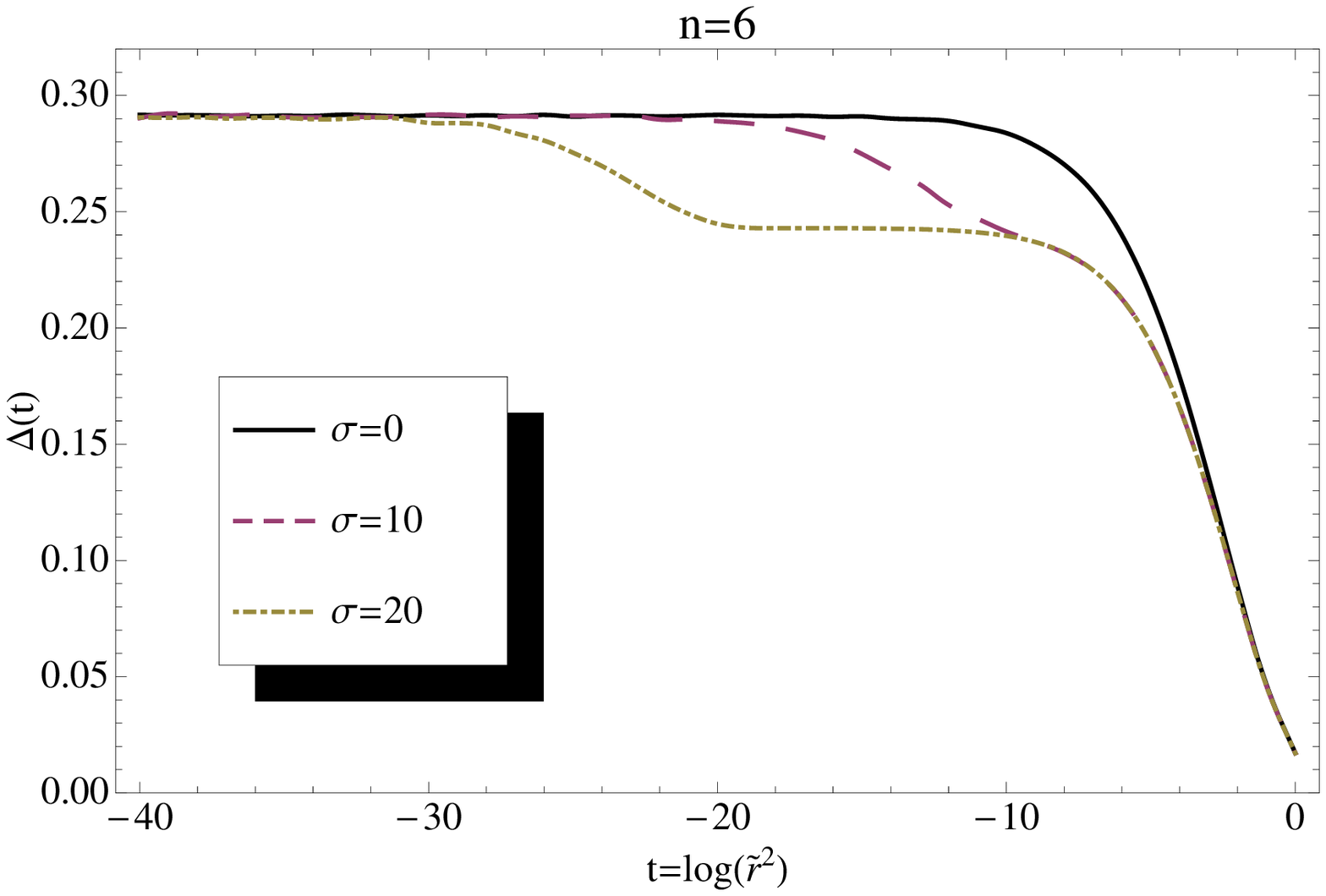}
\includegraphics[width=0.52\textwidth]{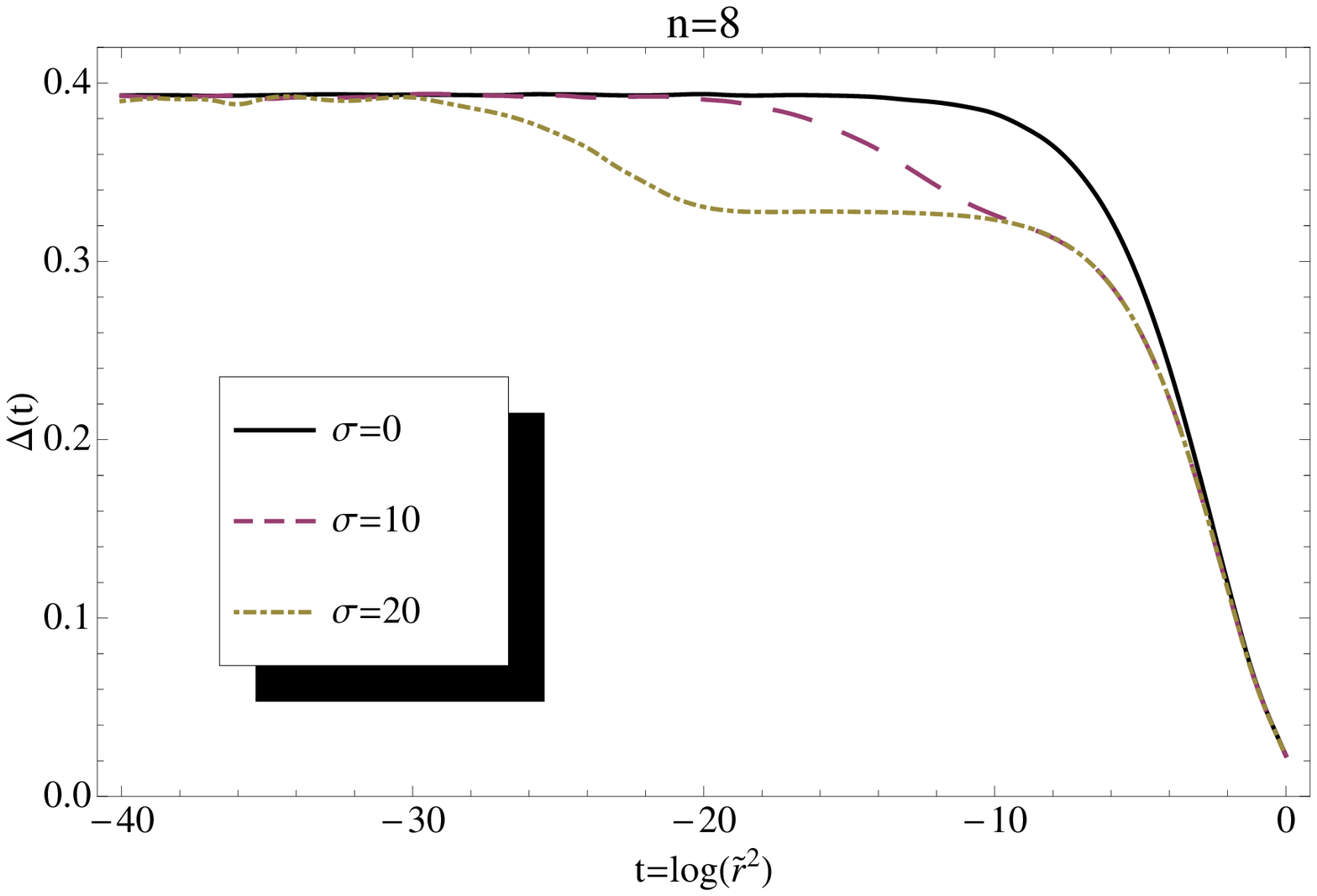}
\label{sggraphics} \caption{The function $\Delta(t):=\Delta^{\mathcal{T}}(t)$ with
$t=2\log(\tilde{r})$ and $\tilde{r}=m r$. In these figures
we show the behaviour of $\Delta^{\mathcal{T}}(t)$ along the renormalization
group flow, from the infrared to the ultraviolet fixed point, for
different values of the resonance parameter $\sigma$. Our results
are consistent with (\ref{known1}) and $c=1$ for the first plateau
and (\ref{known1}) with $c=6/5$ for the second plateau. }
\end{figure}
The four particle contribution is also quite simple to compute, as
only few form factors contribute. This is because, for each copy
of the model, the only non-vanishing four particle form factors of
the energy-momentum tensor are
$F^{\Theta|+-}_{2+2}(\theta_1,\theta_2;\theta_3,\theta_4)$ and all
other form factors that can be obtained from this one by changing
the particle ordering. This form factor was given explicitly in
\cite{CastroAlvaredo:2000em}. Together with our solution
(\ref{fourpar}) and the ansatz (\ref{ansatz}) it gives the four
particle contribution
\begin{equation}
\begin{split}
&\Delta^{\mathcal{T}}_{4}(\tilde{r}_{0}) =-\frac{\cos\left(
\frac{\pi}{2n} \right)^{2}}{256n\pi^{3}e^{\sigma/n}}
\int_{-\infty}^{\infty}d\theta_{1}d\theta_{2}d\theta_{3}d\theta_{4}
\frac{(1+\tilde{r}_{0}(\sum_{i=1}^{4}\cosh(\theta_{i}))
e^{-\tilde{r}_{0}(\sum_{i=1}^{4}\cosh(\theta_{i}))}}{(\sum_{i=1}^{4}\cosh(\theta_{i}))^{2}}
e^{(\theta_{31}+\theta_{42})/2}\\
&
\frac{(2+\sum_{i<j}^{4}\cos(\theta_{ij}))\left[\prod_{i<j}^{4}\left( F_{\text{min}}^{\mathcal{T}|\mu_{i}\mu_{j}}(\theta_{ij}) \right)^{*}
  F_{\text{min}}^{\Theta|\mu_{i}\mu_{j}}(\theta_{ij})\right]Q_{2+2}^{+-}(x_1,x_2;x_3,x_4) e^{-(\theta_1+\theta_2+\theta_3+\theta_4)/n}}{\cosh \left(
\frac{\theta_{12}}{2}  \right) \cosh \left(  \frac{\theta_{34}}{2}
\right) \left( \cosh \left(\frac{\theta_{12}}{n}  \right)-\cos \left(\frac{\pi}{n}
\right)   \right)  \left( \cosh \left(\frac{\theta_{34}}{n}  \right)-\cos \left(\frac{\pi}{n}
\right)   \right)  },
 \end{split}
 \label{4contr}
\end{equation}
where $\mu_1, \mu_2=+$ and $\mu_3,\mu_4=-$.
This contribution, when added to (\ref{2contr}) should bring the
value of $\Delta^{\mathcal{T}}$ closer to the expected one,
which is obtained by setting $c=6/5$ in (\ref{known1}). Our
numerical results shown in Fig.~1 clearly demonstrate this to be
the case for various values of $n$. As $t\rightarrow -\infty$ the
functions $\Delta(t)$ all approach the expected value
(\ref{known1}) for $n=2,4,6$ or 8 and $c=6/5$ with great accuracy.

In Fig.~1 we also see that the function $\Delta(t)$ exhibits two
finite plateaux along the renormalization group flow, which in
numerical terms exactly correspond to the two particle and four
particle contributions. The position at which the second plateau
emerges changes as a function of $\sigma$, as is also illustrated
in the figure. An entirely similar behaviour was found in
\cite{CastroAlvaredo:2000ag} for the $c$-function of the same
model and in \cite{CastroAlvaredo:1999em} for its effective central  charge.
A detailed physical interpretation has been given there.

Unfortunately, the errors on the 6 particles contribution were too
large to give acceptable results.

\section{Conclusions}
This work has been inspired by the relationship between branch
point twist fields and the ground state entanglement entropy of
1+1 dimensional integrable QFTs. This connection provides one of
the main motivations to study the properties of this kind of twist
field. Particularly valuable information is provided by the form
factors. They can be directly employed to generate a low energy
(infrared) form factor expansion of the R\'enyi entropy
(\ref{renyi}) or, as we have seen here, to extract the ultraviolet
conformal dimension of the twist field. From a mathematical
viewpoint one can also consider the equations
(\ref{1})-(\ref{kre}) in their own right, investigate the
properties of their solutions and try to find patterns as the
particle numbers are increased.

In this paper we have constructed higher particle  solutions to
(\ref{1})-(\ref{kre}) for two particular models and have tested
those solutions against the $\Delta$-sum rule. Our computations
have revealed a number of interesting features: first, although
the solution procedure and equations have many similarities with
those for other local fields, it is considerably harder to find
higher particle solutions for the twist field. This is mainly due
to the increased number of poles the form factors have within the
extended physical sheet. As a consequence, even for simple models
it does not seem possible to find the nice closed
determinant formulae found for example in
\cite{FMS,KK,CastroAlvaredo:2000em,CastroAlvaredo:2000nk}.

Second, this extra difficulty becomes particularly  clear for the
Ising model for which one would expect to be able to find more
general results. In fact, the Ising model case suggests that
solving equations (\ref{1})-(\ref{kre}) may not be the most
effective way to construct the form factors of twist fields. In
line with this, we would like to investigate whether or not other
approaches, such as angular quantization
\cite{Lukyanov:1992sc,Lukyanov:1993pn,Brazhnikov:1997wn} may be
more suitable. In particular, the expression (\ref{full}) found in
\cite{entropy} was obtained in a very natural way by the latter
method.

Finally, for the RT-model we have noted that solutions  to the
form factors equations for branch point twist fields are generally
not unique. This lack of uniqueness is not unexpected.  This
is because this geometric picture of the twist field as an object
that connects the various sheets in a Riemann surface is not the
only feature that characterizes the twist field. As its name
indicates it is mainly characterized by a branch cut. One may also
change the features of the twist field by putting other fields at
the corresponding branch point. The expectation is that such changes will give rise to other twist fields
with higher ultraviolet conformal dimensions. Our analysis of the RT-model,
including the investigation of the cluster decomposition property
of form factors, has revealed that some of these other twist
fields correspond  to non primary fields at conformal level and
are likely to be related to composite fields involving the
entropy-related twist field and other fields of the theory. We
have found that for the RT-model and generally any model with a
single particle spectrum, the most general solution for the
$2k$-particle form factor of the twist field depends on $k$ free
parameters.  It would be very interesting to count the number of
twist fields systematically by counting the number of solutions to
(\ref{1})-(\ref{kre}). Also, we would like to investigate the
conformal dimensions of all these extra fields and generally
identify their counterparts at conformal level.

Concerning the numerical computations performed here,  our aim has
been to test the few form factor solutions obtained for two
theories: the roaming trajectories model and the
$SU(3)_2$-homogeneous sine-Gordon model. Both share the appearance
of staircase patterns for the associated effective central charges
\cite{roaming,CastroAlvaredo:1999em}. For the HSG-model the same
pattern has been reproduced for Zamolodchikov's $c$-function
\cite{Zamc,CastroAlvaredo:2000ag} and for the conformal dimensions
of certain local fields \cite{CastroAlvaredo:2000ag}. Our numerics
demonstrate that such pattern is again reproduced for the
conformal dimension of the twist field which exhibits two plateaux
at $\Delta^{\mathcal{T}}=\frac{1}{24}\left(n-\frac{1}{n}\right)$
and $\Delta^{\mathcal{T}}=\frac{1}{20}\left(n-\frac{1}{n}\right)$.
For the RT-model we focused on the first and second steps in the
staircase pattern only, corresponding to
$\Delta^{\mathcal{T}}=\frac{1}{48}\left(n-\frac{1}{n}\right)$ and
$\Delta^{\mathcal{T}}=\frac{7}{240}\left(n-\frac{1}{n}\right)$,
respectively.

An interesting conclusion that  can be drawn from our numerics, specially for the $SU(3)_2$-HSG model, is that the function $\Delta^{\mathcal{T}}(r_0)$ given by (\ref{delta2}) behaves exactly as
\begin{equation}\label{delce}
    \Delta^{\mathcal{T}}(r_0)=\frac{c(r_0)}{24}\left(n-\frac{1}{n} \right),
\end{equation}
where $c(r_0)$ is Zamolodchikov's $c$-function.  Should this identity be exact, it would mean that  the function $\Delta^{\mathcal{T}}(r_0)$ is positive definite and monotonically decreasing (as a function of $t=2\log(mr_0)$), just as $c(r_0)$. It is however not obvious why (\ref{delta2}) should have these features. Clearly, they tell us something fundamental about the nature of the correlation function $\langle \Theta(r) \mathcal{T}(0)\rangle$ and the branch point twist field. It would be very interesting to investigate this further, particularly its implications (if any) for the bi-partite entanglement entropy of integrable QFTs \cite{new}.
\paragraph{Acknowledgments:}
The authors are grateful to Patrick Dorey, Benjamin Doyon and Andreas Fring for their feedback on this manuscript.

\appendix
\section{Explicit formulae for $Q_4(x_1,x_2,x_3,x_4)$ and $K_4(x_1,x_2,x_3,x_4)$}
The constants in (\ref{q4}) are given by
\begin{eqnarray}
  \gamma &=& 2\left( 1 + 2\cos (\frac{\pi }{n}) \right) \sec (\frac{\pi }{2n})
  \sin (\frac{\left(B-4 \right) \pi }{4n})
  \sin (\frac{\left( 2 + B \right) \pi }{4n}),\\
  \delta &= &-\left(4\cos (\frac{\pi }{2n}) - \cos (\frac{3\pi }{2n}) -
    \cos (\frac{\left(B-1 \right) \pi }{2n}) \right) \sec (\frac{\pi
    }{2n}),\\
    \eta &=& 3 + 2\cos (\frac{\pi }{n}) +
  4\cos (\frac{\pi }{2n})
   \cos (\frac{\left( B-1 \right) \pi }{2n}),\\
   \xi &=&2\left( 1 + 3\cos (\frac{\pi }{n}) - \cos (\frac{2\pi }{n}) +
    8{\cos (\frac{\pi }{2n})}^3
     \cos (\frac{\left( B-1 \right) \pi }{2n}) +
    \cos (\frac{\left( B-1 \right) \pi }{n}) \right),\\
    \lambda &=& -2\left( 6 + 6\cos (\frac{\pi }{n}) + 4\cos (\frac{2\pi }{n}) +
    \cos (\frac{3\pi }{n})+
    \left( 1 + 2\cos (\frac{\pi }{n}) \right)
     \cos (\frac{\left(B-1 \right) \pi }{n})\right)\nonumber\\ && -
    4\left( 5\cos (\frac{\pi }{2n}) +
       2\cos (\frac{3\pi }{2n}) + \cos (\frac{5\pi }{2n})
       \right) \cos (\frac{\left( B-1 \right) \pi }{2n}),\\
     \rho &=&8{\cos (\frac{\pi }{n})}^2
  \left( 3 + 3\cos (\frac{\pi }{n}) + \cos (\frac{2\pi }{n}) +
    8{\cos (\frac{\pi }{2n})}^3
     \cos (\frac{\left( B-1 \right) \pi }{2n})  \right)\nonumber\\
     && + 8{\cos (\frac{\pi }{n})}^2
    \left( 1 + 2\cos (\frac{\pi }{n}) \right)
     \cos (\frac{\left( B-1 \right) \pi }{n}) .
\end{eqnarray}
All the constants above are real for $B$ real and they remain real
when $B=1-\frac{2i\theta_0}{\pi}$, as one would expect.

The constants in (\ref{k4}) are given by
\begin{eqnarray}
  A &=& -\frac{1}{\left( 1 + 2\cos (\frac{\pi }{n}) \right)^3}, \\
  B&=& \frac{2\left( 1 + \cos (\frac{\pi }{n}) \right) }
  {{\left( 1 + 2\cos (\frac{\pi }{n}) \right) }^3},\\
  C &=&-\frac{2\left( 2 + \cos (\frac{\pi }{n}) \right) }
  {{\left( 1 + 2\cos (\frac{\pi }{n}) \right) }^2},\\
  D &=& -\frac{16\cos^4 (\frac{\pi }{2n})}{{\left( 1 + 2\cos (\frac{\pi }{n}) \right) }^3}, \\
  E &=& \frac{8{\cos (\frac{\pi }{2n})}^2
    \left( 3 + 6\cos (\frac{\pi }{n}) + \cos (\frac{2\pi }{n}) \right) }{{\left( 1 +
       2\cos (\frac{\pi }{n}) \right) }^3}, \end{eqnarray}
    \begin{eqnarray}
  F&=& \frac{2\left( 2\cos (\frac{\pi }{n}) + \cos (\frac{2\pi }{n}) \right) }
  {{\left( 1 + 2\cos (\frac{\pi }{n}) \right) }^3},\\
       G&=&-\frac{16{\cos (\frac{\pi }{2n})}^2\cos (\frac{\pi }{n})\,
    \left( 2 + \cos (\frac{\pi }{n}) \right) }{{\left( 1 + 2\cos (\frac{\pi }{n}) \right) }^
    2},\\
     H &=& \frac{128{\cos (\frac{\pi }{2n})}^4{\cos (\frac{\pi }{n})}^2}
  {{\left( 1 + 2\cos (\frac{\pi }{n}) \right) }^3},\\
  I &=& \frac{8{\left( \cos (\frac{\pi }{2n}) + \cos (\frac{3\pi }{2n}) \right) }^2
    \left( 3 + 2\cos (\frac{\pi }{n}) + \cos (\frac{2\pi }{n}) \right) }{{\left( 1 +
       2\cos (\frac{\pi }{n}) \right) }^3},\\
       J&=&- \frac{256 {\cos (\frac{\pi }{2n})}^4{\cos (\frac{\pi }{n})}^4}
  {{\left( 1 + 2\cos (\frac{\pi }{n}) \right) }^3}.
\end{eqnarray}


\end{document}